\documentclass[%
 reprint,longbibliography,preprintnumbers,
nofootinbib,
 amsmath,amssymb,
 aps,
 pre,
]{revtex4-1}
\pdfoutput=1
\usepackage{graphicx}
\usepackage[utf8]{inputenc}
\usepackage{flushend}
\usepackage{dcolumn}
\usepackage{bm}
\usepackage{balance}
\usepackage{mathrsfs,epigraph,bm}

\usepackage{tikz}
\usetikzlibrary{mindmap,trees}

\usepackage{tcolorbox}
\usepackage{tabularx}
\usepackage{array}
\usepackage{colortbl}
\tcbuselibrary{skins}

\usepackage{tcolorbox}
\definecolor{mycolor}{rgb}{0.122, 0.435, 0.698}
\makeatletter
\newcommand{\mybox}[1]{%
  \setbox0=\hbox{#1}%
  \setlength{\@tempdima}{\dimexpr\wd0+13pt}%
  \begin{tcolorbox}[colframe=mycolor,boxrule=0.5pt,arc=4pt,
      left=6pt,right=6pt,top=6pt,bottom=6pt,boxsep=0pt,width=0.98\linewidth]
    #1
  \end{tcolorbox}
}
\makeatother

\newtcbox{\boxbello}{on line,
  colframe=orange,colback=yellow!20!white,
  boxrule=0.5pt,arc=4pt,boxsep=0pt,left=6pt,right=6pt,top=6pt,bottom=6pt,width=0.98\linewidth}
\newcolumntype{Y}{>{\raggedleft\arraybackslash}X}

\tcbset{tab1/.style={fonttitle=\bfseries\large,fontupper=\normalsize\sffamily,
colback=yellow!10!white,colframe=red!75!black,colbacktitle=purple!40!white,
coltitle=black,center title,freelance,frame code={
\foreach \n in {north east,north west,south east,south west}
{\path [fill=purple!75!black] (interior.\n) circle (3mm); };},}}

\tcbset{tab2/.style={enhanced,fonttitle=\bfseries,fontupper=\normalsize\sffamily,
colback=yellow!10!white,colframe=red!50!black,colbacktitle=purple!40!white,
coltitle=black,center title}}
\usepackage[normalem]{ulem}

\usepackage[colorlinks = true,
            linkcolor = blue,
            urlcolor  = blue,
            citecolor =green,
            anchorcolor = blue]{hyperref}
\usepackage{verbatim}
\usepackage{color,ulem}
\usepackage[english]{babel}

\usepackage[utf8]{inputenc}
\input Starburst.fd
\newcommand*\initfamily{\usefont{U}{Starburst}{xl}{n}}\initfamily

\newcommand{\beq}{\begin{eqnarray}}
\newcommand{\eeq}{\end{eqnarray}}
\usepackage{amsmath}
\usepackage{tikz}
\usetikzlibrary{decorations.pathmorphing}
\usetikzlibrary{shapes.misc}
\tikzset{cross/.style={cross out, draw=black, minimum size=8*(#1-\pgflinewidth), inner sep=0pt, outer sep=0pt},
cross/.default={1pt}}
\usetikzlibrary{patterns,math}
\begin{document}

\title{Deformations, relaxation and broken symmetries in liquids, solids and glasses:\\ a unified topological field theory}

\author{\textbf{Matteo Baggioli}$^{1,2}$}%
 \email{b.matteo@sjtu.edu.cn}
 \author{\textbf{Michael Landry}$^{3}$}
 \email{ml2999@columbia.edu}
\author{\textbf{Alessio Zaccone}$^{4,5}$}%
 \email{alessio.zaccone@unimi.it}
 
 \vspace{1cm}
 
\affiliation{$^{1}$Wilczek Quantum Center, School of Physics and Astronomy, Shanghai Jiao Tong University, Shanghai 200240, China}
\affiliation{$^{2}$Shanghai Research Center for Quantum Sciences, Shanghai 201315.}
\affiliation{$^{3}$Department of Physics, Center for Theoretical Physics,
Columbia University, 538W 120th Street, New York, NY, 10027, USA.}
\affiliation{$^{4}$Department of Physics ``A. Pontremoli'', University of Milan, via Celoria 16,
20133 Milan, Italy.}
\affiliation{$^{5}$Cavendish Laboratory, University of Cambridge, JJ Thomson
Avenue, CB30HE Cambridge, U.K.}

\begin{abstract}
We combine hydrodynamic and field theoretic methods to develop a general theory of phonons as Goldstone bosons in crystals, glasses and liquids based on non-affine displacements and the consequent Goldstones phase relaxation. We relate the conservation, or lack thereof, of specific higher-form currents with properties of the underlying deformation field -- non-affinity -- which dictates how molecules move under an applied stress or deformation. In particular, the single-valuedness of the deformation field is associated with conservation of higher-form charges that count the number of topological defects. Our formalism predicts, from first principles, the presence of propagating shear waves above a critical wave-vector in liquids, thus giving the first formal derivation of the phenomenon in terms of fundamental symmetries. The same picture provides also a theoretical explanation of the corresponding ``positive sound dispersion'' phenomenon for longitudinal sound. Importantly, accordingly to our theory, the main collective relaxation timescale of a liquid or a glass (known as the $\alpha$ relaxation for the latter) is given by the phase relaxation time, which is not necessarily related to the Maxwell time. Finally, we build a non-equilibrium effective action using the in-in formalism defined on the Schwinger-Keldysh contour, that further supports the emerging picture. In summary, our work suggests that the fundamental difference between solids, fluids and glasses has to be identified with the associated generalized higher-from global symmetries and their topological structure, and that the Burgers vector for the displacement fields serves as a suitable topological order parameter distinguishing the solid (ordered) phase and the amorphous ones (fluids, glasses).
\end{abstract}

\maketitle
\section{Introduction}

\begin{figure}[ht]
    \centering
    \includegraphics[width=0.9\linewidth]{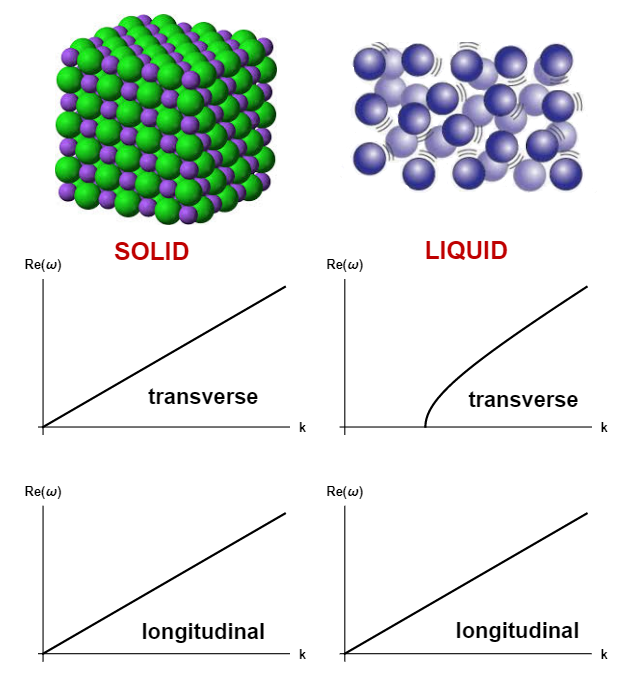}
    \caption{The difference between solids and liquids from the point of view of the collective excitations in the transverse and longitudinal sectors.}
    \label{fig:prima}
\end{figure}

From a microscopic viewpoint, solids, liquids and glasses are profoundly different. The short-scale dynamics are strongly dependent on their structural differences, which are revealed in several physical observables such as the vibrational density of states $g(\omega)$ and the specific heat $c_v(T)$. Solids are well-described by Debye theory \cite{kittel2004introduction,chaikin2000principles}, which predicts that $g(\omega)\sim \omega^2$ and $c_v \sim T^3$. Liquids on the contrary, yield $g(\omega)\sim \omega$ \cite{Berne,Douglas,zaccone2021universal} and a specific heat which decreases monotonically with temperature \cite{GRANATO2002376,PhysRevE.57.1717,Bolmatov2012}. Glasses are even more mysterious since they display an anomalous peak (Boson peak) in both the density of states and the specific heat \cite{PhysRevB.4.2029,Shintani2008,PhysRevB.87.134203,PhysRevLett.96.045502}, and exhibit an unusual linear relationship $c_v\sim T$ at low temperature \cite{PhysRevB.4.2029}. 
These macroscopically different behaviors arise from the microscopic idiosyncrasies of solids, liquids and glasses. 
In particular, although solids, liquids and glasses feature propagating longitudinal sound waves (albeit with difference velocities), the dynamics of transverse shear waves are very different. Solids exhibit propagating shear waves down to arbitrarily low momentum and their speed is set by the shear elastic modulus $G$. Liquids, on the contrary, do not display propagating shear waves at low momenta and their low-momentum dynamics are governed by a shear diffusive mode, which is simply the manifestation of momentum conservation. Nevertheless, propagating shear waves appear above a certain critical momentum, which has historically been called the Frenkel theory of liquids, or more recently, the \textit{k-gap} theory\cite{BAGGIOLI20201} (see Fig.\ref{fig:prima}).\\

The existence of this feature in liquids is supported by numerous molecular dynamic (MD) simulations \cite{PhysRevLett.92.065001,PhysRevLett.85.2514,PhysRevLett.84.6026,PhysRevLett.97.115001,Hosokawa_2015,doi:10.1063/1.5088141,PhysRevLett.125.125501,PhysRevLett.118.215502,PhysRevB.101.214312,doi:10.1063/1.5050708,PhysRevE.85.066401} and it might be the fundamental mechanism behind the experimental observation of solid-like behaviours in confined liquids \cite{Noirez_2012,Kume2020,kume2020unexpected}. The starting -- or as we will see, ending -- point of the Frenkel or \textit{k-gap theory} is the simple equation
\begin{equation}
    \omega^2\,+\,i\,\omega\,\Gamma\,=\,v^2\,k^2\,, \label{wrong}
\end{equation}
which is known as \textit{telegraph equation}, originally introduced by Heaviside \cite{heaviside_oliver_1876_1645475}. Here $\Gamma$ is the relaxation rate, the inverse of the relaxation time $\tau$.
This equation is simply the Fourier transform of the following one-dimensional manipulation of the Navier-Stokes equation developed by Y. Frenkel~\cite{frenkel} to accommodate viscoelasticity à la Maxwell
\begin{equation}
\eta\, \partial_{x}^{2} v=\rho\, \tau \partial_{t}^{2} v + \rho \,\partial_{t} v,
\nonumber
\end{equation}
where $v$ is the fluid velocity component in either the $y$ or $z$ direction, $\tau=\eta/G_{\infty}$ is the Maxwell viscoelastic relaxation time, $G_{\infty}$ is the infinite-frequency shear modulus, $\eta$ is the kinematic viscosity and $\rho$ is the density.
Contrary to the standard Navier-Stokes equation, the above hydrodynamic equation contains a second derivative in time, thus allowing for sound propagation with speed $c_{s}=\sqrt{G_{\infty}/\rho}$. This sound mode is a result of implementing Maxwell's so-called ``viscoelastic interpolation'' inside the Navier-Stokes equation. As a matter of fact, the above equation is a wave equation with damping. Unfortunately, Maxwell's interpolation with Frenkel's suggestion is not sustained by any formal and consistent theory. It is an ad-hoc phenomenological procedure. More precisely, it is not even clear if Frenkel's derivation of Eq.\eqref{wrong} is compatible with conservation laws and a well-defined hydrodynamic framework\footnote{As we will show, the only way to achieve this dynamics for the collective shear waves is by adding \textit{phase relaxation}, which was not present in the original framework of Maxwell and Frenkel.}.\\
The same equation can be derived in the context of \textit{generalized hydrodynamics} \cite{MOUNTAIN1977225,boon1980molecular} but, again, a mathematical justification of this framework from fundamental principles (e.g. symmetries) is still absent\footnote{Generalized hydrodynamics simply assume that the thermodynamics and transport coefficients become $k,\omega$ dependent to reflect the microscopic details of the molecular environment without modifying the hydrodynamic equations.}. See \cite{hansen2006theory} for a classical textbook discussing all these issues.\\

This problem has produced a large amount of confusion regarding the nature of the relaxation rate $\Gamma$, which is fundamental in the phenomenological applications of this model. In particular, the Maxwell-Frenkel idea of taking $\Gamma$ as the shear relaxation rate $\Gamma=G_{\infty}/\eta$ is quite disputable and not compatible with hydrodynamics. It is indeed well-known that the shear viscosity $\eta$, combined with the high-frequency shear elasticity modulus $G_{\infty}$, does not induce collective dynamics for transverse shear waves as in Eq.\eqref{wrong} (see Landau textbook \cite{Landau_elasticity} or Chaikin and Lubensky's book \cite{Chaikin}). On the contrary, it gives attenuated transverse sound waves of the form
\begin{equation}
    \omega_{\textit{T}}\,=\,\pm\,v\,k\,-\,i\,\frac{\Gamma_T}{2}\,k^2\,+\,\mathcal{O}(k^3)\,,\qquad v^2\,=\,\frac{G}{\rho}\,,\,\,\,\,\Gamma_T\,=\,\frac{\eta}{\rho}\,,
\end{equation}
which are typical of viscoelastic systems \cite{Chaikin,Delacretaz:2017zxd,Armas:2019sbe}, where $\rho$ is the mass density of the medium.

At the same time, it is obvious that the relaxation rate $\Gamma$ in Eq.\eqref{wrong} cannot come in any way from the explicit breaking of spacetime symmetries, since a non-conserved momentum would also destroy the longitudinal propagating sound.

Finally, rotations play no role in the Goldstones counting (contrary to what was suggested in \cite{Bolmatov2013}) since it is well-known that they are not independent uniform symmetries and they are always ``subsumed'' by translations. Indeed, in standard material media, or standard spacetime, the Lie generators of rotations are just linear combinations of the generators of translations, and do not generate additional Goldstone gapless modes, a fact that was already pointed out by Mermin~\cite{Mermin} (and in its modern formulation is due to the so-called \textit{inverse Higgs constraint} \cite{Brauner:2014aha}). Different is the case in liquid crystals and other Cosserat media~\cite{Terentjev,Maugin}.\\ 

This entire discussion thus reduces to a fundamental question: \textit{what is the difference between liquids and solids at the level of fundamental symmetries} ?\\

The purpose of this work is to provide an answer to this question, and more precisely {\it what} is the relaxation rate $\Gamma$ appearing in Eq.\eqref{wrong}. Importantly, this is not a purely academic question but it plays a fundamental role in determining the validity of the current theory of the liquid state.\\

A partial answer to this question has been put forward in \cite{Nicolis:2013lma,Nicolis:2015sra} by noticing that fluids enjoy more symmetries than solids, and in particular they are invariant (within the linear approximation) under \textit{volume preserving diffeomorphisms} -- transformations of the fluid world-volume that do not change the total volume of the system. It is easy to show that invariance under this group constrains the static (equilibrium) shear elastic modulus $G$ to be zero. However, connecting this internal symmetry with the microscopic dynamics and physical features is very hard. Moreover, this does not explain the appearance of propagating phonons in liquids, and thus does not account for a finite elastic modulus above a certain critical momentum.\\

A closely related question is: what is the fate of the transverse phonons upon transitioning from liquids to solids? In particular, what happens upon transitioning from the liquid state, with one ungapped longitudinal phonon and heavily gapped transverse phonons, to the amorphous solid state (glasses), with two ungapped (or apparently so \cite{Biroli2016}) transverse phonons, in addition to the usual ungapped longitudinal phonon? Evidently, these questions touch on the unsolved problem of the glass transition \cite{Biroli2016,Berthier_review}, i.e. how does a liquid turn into a solid glass without an apparent change of microscopic structure (aside from the apparent slow-down of the atomic/molecular dynamics)? A deeply related question is therefore: what changes in the underlying fundamental symmetries (e.g. at the level of Goldstone bosons) accompany this transition? \\

In this work, we take a perspective based on symmetries, hydrodynamics and classical field theory. 
Although the mathematical theory underlying the hydrodynamic description of liquids has been studied for quite some time,
the results of these studies have not been
collected in one place to provide a clear and concise guide to both theorists and experimentalists who might use them.

In this largely expository paper, using different methods, the dynamics of a liquid will be explained by considering a system that spontaneously breaks translations along with the presence of the so-called \textit{phase relaxation}~\cite{Grozdanov:2017kyl}. The phase relaxation, whose meaning will be explained in the next sections, is the key to understanding the disappearance of propagating shear waves at small momentum in liquids. It is induced microscopically by the presence of multi-valued, non-affine displacements in the deformation field of liquids and it can be connected to some sort of topological properties of the material using the framework of generalized global symmetries \cite{Gaiotto:2014kfa}. The same process  happens to a \textit{quantitatively} lesser extent in glasses, where weaker phase relaxation induces a partial or apparent restoring of the transverse phonons.\\

In a sense, our work suggests that the fundamental distinction between solids, liquids and glasses must be found at the level of higher-form generalized global symmetries and their associated topological structure. Additionally, the Burgers vector measured in the displacement field under deformation can efficiently serve as a suitable order parameter for distinguishing the solid (ordered) phase from the amorphous ones (fluids, glasses). The Burgers vector is indeed zero in the ordered solid phase (with no defects) and it acquires finite values in the glassy or fluid phases.\\

In summary, from first principles, we construct a fully consistent field theory of liquids and solids based on symmetries, which agrees with all known experimental observations. More specifically, it explains both the absence of propagating shear waves at small momentum and their presence at higher momentum in liquids, and it connects this feature to their fundamental non-affine dynamics of deformation. Moreover, it presents a new understanding of the glass transition as the point at which the phase relaxation time becomes large enough such that the k-gap disappears and transverse waves are again propagating at all scales. Our results are backed up by several different methods: (I) a hydrodynamic description, (II) a field theory formulation in terms of higher form symmetries, (III) a differential-geometry study based on the deformation compatibility constraint and (IV) a formal non-equilibrium effective action defined by using the in-in formalism on the Schwinger-Keldysh (SK) contour.\\

For completeness, let us emphasize that the questions and the phenomena described in this paper are well known in the physics of liquids. Previous theoretical frameworks include viscoelastic theory \cite{PhysRevLett.53.401}, generalized hydrodynamics \cite{PhysRevA.38.271,PhysRevA.22.287}, mode coupling theory \cite{gotze1995mode}, self-consistent relaxation theory \cite{mokshin2021quasi} and “k-gap” theory \cite{BAGGIOLI20201}. Importantly, all these frameworks are in a sense phenomenological and not derived from fundamental principles. In this aspect, our approach is different and in a way complementary to those mentioned above.

The paper is structured as follows.
In Section \ref{sec2} we present a description of non-affine displacements in liquids and glasses and we demonstrate that they can be described in terms of a Burgers circuit and a Burgers vector, in analogy with dislocations in crystals: this reveals the topological multi-valuedness of the displacement field in disordered systems; in Section \ref{sec3} we briefly summarize the main ideas behind the k-gap theory of liquids; in \ref{sec4} we provide the hydrodynamic description of systems with spontaneously broken translations and phase relaxation; in Section \ref{sec5}
we discuss how phase relaxation can be understood in terms of a higher-form global symmetry; in Section \ref{sec6} we connect the non-affine deformations of liquids and glasses with phase relaxation; in Section \ref{secnew} we compare our formalism with the Maxwell-Frenkel theory; in Section \ref{secnew} we provide explicit and concise proofs of all the main statements described in the previous parts; in Section \ref{sec7} we build a Schwinger-Keldysh action principle for systems with phase relaxation and we provide a comprehensive Goldstone bosons counting for liquids, glasses and crystals; in Section \ref{secglass} we briefly discuss the implications of our theory for glasses and finally in Section \ref{sec8} we conclude with some comments and future directions.
\section{Non-affine displacement field in liquids and glasses}\label{sec2}
\subsection{Non-affine deformations in amorphous media: mechanism and phenomenology}
\begin{figure}[ht]
    \centering
    \includegraphics[width=0.95\linewidth]{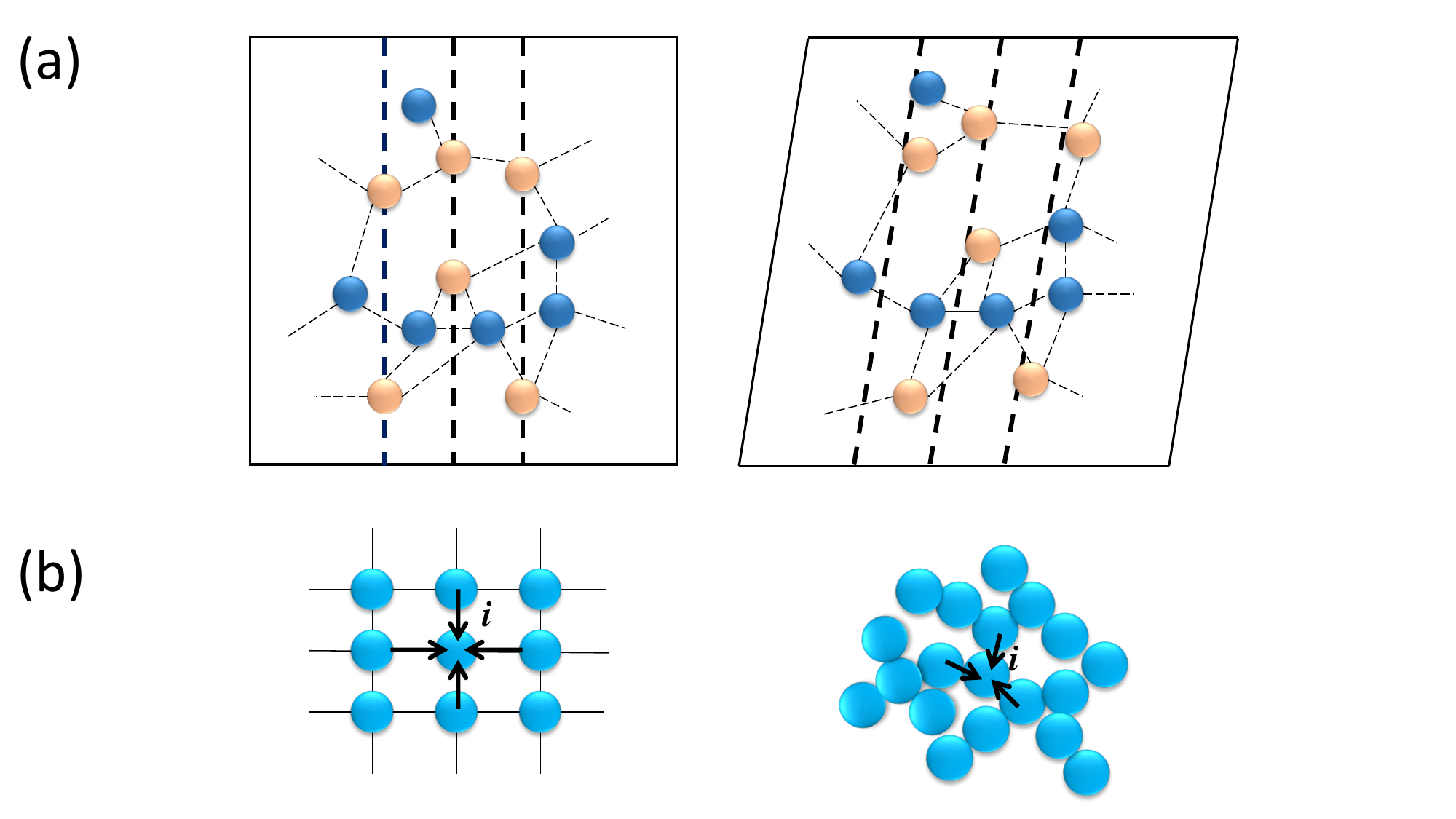}
    \caption{Schematic illustration of nonaffine displacements in amorphous media. \textbf{Panel (a)} shows the rearrangements or displacements of atoms upon application of an external shear strain. If the deformation were affine, atoms which  sit exactly on the dashed lines in the underformed frame (left) would still sit exactly on dashed lines also in the deformed frame (right). However, in a disordered environment this does not happen: the atoms that were sitting on the dashed lines in the undeformed frame are no longer sitting on the dashed lines in the deformed frame, but are displaced from them. The distance from the actual positions of the atoms to the dashed line corresponds to the non-affine displacements. \textbf{Panel (b)} provides a visual explanation of the origin of non-affine displacements in disordered environments. Left figure shows a perfect lattice where, upon applying a small deformation, the nearest-neighbour forces from surrounding atoms cancel each other out in the affine positions, so there is no need for non-affine displacements to arise. In the right figure, instead, the tagged atom $i$ is not a center of inversion symmetry, which implies that nearest-neighbour forces from surrounding atoms do not balance in the affine position, hence a net force arises which triggers the non-affine displacement in order to maintain mechanical equilibrium.}
    \label{fig:zero}
\end{figure}

\begin{figure}[ht]
    \centering
      \includegraphics[width=0.6\linewidth]{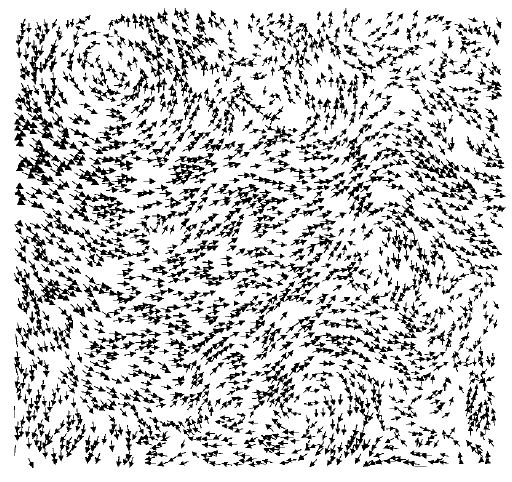}
    \caption{Non-affine displacements resulting from a shear in the $x,y,$ plane. Notice the vortex-like pattern. Figure reproduced with permission of the American Physical Society from Ref.\cite{PhysRevE.72.066619}.}
    \label{fig:liquids2}
\end{figure}

In liquids, glasses and granular-jammed systems, the physics of deformations at the microscopic (atomic, molecular, or particle) level is very different from that of perfect crystals at low temperature. Under an infinitesimal constant shear, the atoms of each of these systems will move and settle into new equilibrium positions. For a crystalline solid, it is natural to suppose that the new positions of the atoms be related to the old positions by a simple affine transformation that can be calculated from the strain tensor; for disordered systems, however, no such affine transformation exists. In particular, because of the disorder, a reference atom cannot be at mechanical equilibrium in the spatial position prescribed by such an affine transformation \cite{Lemaitre,Zaccone_2011,doi:10.1142/S0217984913300020,Zaccone_review}. As a result, the forces transmitted by the nearest neighbour atoms (on their way towards their respective affine positions) to the reference atom in its affine position do not cancel. In a perfect centrosymmetric crystal, inversion symmetry of the lattice requires that these microscopic forces cancel with each other, but no such symmetry exists for disordered materials like liquids, glasses and granulars, or in a crystal near defects.

Because there is a net force acting on each atom in its affine position, an extra (non-affine) displacement away from the affine position is required in order to maintain the mechanical equilibrium throughout the deformation. The non-affine displacements provide an additional contribution to the standard displacement field, $u_{i}$, of elasticity theory. The non-affine part of the displacement field is random (due to the microscopic disorder) and does not possess any particular symmetry. The mechanism of non-affine deformation is summarized in Fig.\ref{fig:zero}.\\

In general, we can express the displacement of a reference atom in a disordered environment (glass or liquid) by
\begin{equation}
u_{i}(\mathbf{x})=F_{ij}x_{j}+u'_{i}(\mathbf{x})
\label{nonaff_displacement}
\end{equation}
where $F_{ij}$ denotes the components of the deformation gradient tensor, while $u'_{i}(\mathbf{x})$ is the non-affine displacement field. As usual, the deformation gradient tensor can be expressed as $F_{ij}=\delta_{ij}+\gamma_{ij}$, where $\delta_{ij}$ is the Kronecker delta. 
For an externally imposed uniform deformation, the $\gamma_{ij}$ are constants, i.e. independent of $\mathbf{x}$, unlike the non-affine component $u'_{i}(\mathbf{x})$, which varies in an unconstrained manner throughout the material. 

A typical snapshot of the non-affine displacement field $u'_{i}$ calculated numerically in a disordered solid (from \cite{PhysRevE.72.066619}) is shown in Fig.\ref{fig:liquids2}. 
In spite of the evident randomness and absence of particular symmetries of $u'_{i}(\mathbf{x})$, the correlation functions of $u'_{i}(\mathbf{x})$ can be obtained in a semi-deterministic way as shown in Ref.\cite{PhysRevE.72.066619}. In particular, vortex-like patterns are systematically found in the numerically computed $u'_{i}(\mathbf{x})$ in glasses, see also~\cite{Tanguy,Lemaitre,Maloney_2006,Goldenberg_2007}. The same type of phenomenology is observed also for supercooled liquids above the glass transitions, see e.g. Ref.\cite{DelGado}.

\subsection{Microscopic elasticity and softening due to non-affine deformations in disordered solids}
The analysis of the non-affine displacements and their mechanism, is the starting point for the modern microscopic theory of elastic and viscoelastic moduli of materials~\cite{Lemaitre,Zaccone_2011,Palyulin}. 

As shown in previous works, the equation of motion of atom $i$ in a disordered medium (liquid or glass) subjected to an external strain, in mass-rescaled coordinates, can be written for an atom labeled as $\alpha$ as follows\cite{Lemaitre,Palyulin}:
\begin{equation}
\frac{d^2x_{i}^{\alpha}}{dt^2}+\nu\frac{dx_i^{\alpha}}{dt}+H_{ij}^{\alpha\beta}x_j^{\beta}
=\Xi_{i}^{\alpha,kl}\eta_{kl}
\label{Newton}
\end{equation}
where $\eta_{ij}$ is the Green-Saint Venant strain tensor and $\nu$ is a microscopic friction coefficient which arises from long-range dynamical coupling between atoms mediated by anharmonicity of the pair potential~\cite{Zwanzig}. $H_{ij}^{\alpha\beta}$ is the Hessian matrix. Latin indices denote spatial components, Greek indices are atoms labels and the summation convention is implied. The term on the r.h.s. physically represents the effect of the disordered (non-centrosymmetric) environment leading to nonaffine motions (see Fig.\ref{fig:zero}). The equation of motion \eqref{Newton} can also be derived from first principles, from a model particle-bath Hamiltonian as shown in previous work~\cite{Palyulin}. 

Using standard manipulations (Fourier transformation and eigenmode decomposition from time to eigenfrequency~\cite{Lemaitre}), and applying the definition of mechanical stress, we obtain the following expression for the
viscoelastic (complex) elastic constants\cite{Lemaitre,Palyulin}:
\begin{equation}
C_{ijkl}(\omega)=C_{ijkl}^{\textit{Born}}-
\frac{1}{V}\sum_n\frac{\hat{\Xi}_{n,ij}\hat{\Xi}_{n,kl}}{\omega_{p,n}^2-\omega^2+i\omega\nu} \label{nonaffine}
\end{equation}
where $C_{ijkl}^{\textit{Born}}$ is the Born or affine part of the elastic constant tensor, i.e. what survives in the high-frequency limit. Here, $\omega$ represents the oscillation frequency of the external strain field, whereas $\omega_p$ denotes the internal eigenmode frequency of the material. Upon taking the limit $\omega = 0$, the static (zero-frequency or zero deformation-rate) elastic constants can be calculated.

The second negative term on the r.h.s. represents the softening contribution from nonaffine displacements, which reduces the elastic constants of the system, and which also play a dominant role at the plastic yielding transition~\cite{Schall}. The reduction is especially important for the shear modulus, whereas for the bulk modulus the non-affine correction is much smaller for geometric reasons due to the local excluded-volume packing constraints~\cite{Ellenbroek,TerentjevJAP,Schlegel}.

The above formalism has been used to solve the problem of the elasticity of random sphere packings in generic $d$ dimensions in Ref.\cite{Zaccone_2011}, leading to an analytical expressions for the shear modulus $G$:
\begin{equation}
G=\frac{1}{30}\frac{N}{V}\,\kappa\,  R_{0}^{2}\,(z-2d) 
\label{jamming}
\end{equation}
where $N/V$ is the density of spheres in the packing, $\kappa$ is the spring constant of the particle-particle interaction, $R_{0}$ is the distance between two nearest-neighbour particles, and $z$ is the average nearest-neighbour number in the packing. The negative term due to non-affine displacements causes the shear modulus to become zero, $G=0$ when $z=2d$, thus recovering the well known Maxwell rigidity criterion. The above Eq.\eqref{jamming} is in excellent agreement with MD simulations of jammed sphere packings~\cite{O'Hern}, including the numerical prefactor.

The non-affine lattice dynamics framework also provides an adequate description of the low-frequency elasticity of confined liquids, and it predicts the correct law $G \sim L^{-3}$ for the dependence of the low-frequency shear modulus of confined liquids upon the confinement length $L$ as shown in \cite{Zaccone_PNAS_2020,phillips2020universal}, in agreement with experimental observations~\cite{Noirez_2012,https://doi.org/10.1002/pi.2625}.

For bulk liquids, the non-affine deformation formalism combined with equilibrium statistical mechanics (as appropriate for equilibrium fluids), predicts $G=0$ for the zero-frequency shear modulus, as shown in Ref.~\cite{Wittmer} using the stress-fluctuation version of the non-affine formalism valid for quasi-static response at zero deformation rate/frequency. In Ref.~\cite{MizunoPRE}, it has been shown that the stress-fluctuation version of non-affine deformations and the non-affine response formalism discussed here do coincide in the limit of $\omega \rightarrow 0$.

\subsection{Burgers vector for non-affine deformations viewed as topological invariant}

Non-affinity of the displacement field arises due to breaking of point-group inversion symmetry, as explained in the previous sections, and is a common feature of crystals with point defects (interstitials and vacancies~\cite{Dederichs,Milkus,Krausser}), and of glasses and liquids as well. Of course, the breaking of point-group inversion symmetry occurs also in crystals with topological defects, and this is immediately evident upon looking at the atoms that sit of the edge of a dislocation line in  a crystal lattice. Those atoms are obviously not centers of symmetry.

\begin{figure}[ht]
    \centering
    \includegraphics[width=0.85\linewidth]{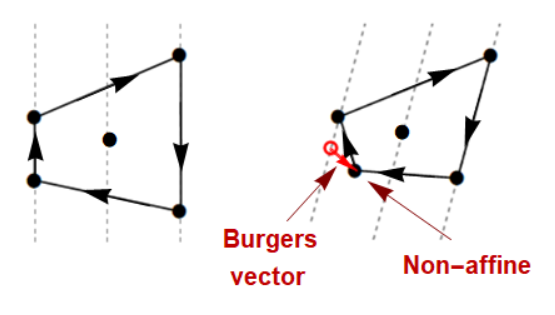}
    \caption{Geometric construction, known as the \textit{Burgers circuit} \cite{chaikin2000principles}, of the topological singularity associated with the non-affine displacement field, leading to the Burgers vector $-b$, around a reference atom in an amorphous solid. For simplicity, and without loss of generality, we consider the case of the left bottom atom being undergoing a non-affine displacement, while the other atoms move affinely.
    The red circle indicates the affine position where the left-bottom atom would have landed, had it been displaced affinely. The red line, which measures the non-affine displacement $u'_{i}$, as defined in Eq.\eqref{nonaff_displacement}, is the \textit{Burgers vector} for the non-affine displacement field in an amorphous lattice.}
    \label{fig:liquids3}
\end{figure}

Therefore, it would be natural to think of deep analogies in the mathematical description of the non-affine deformation field in amorphous media and that of the displacement field around dislocations. It is surprising that these analogies have not been addressed in the previous literature yet, aside from a series of studies which addressed the issue of identifying ``defects'', including dislocations, in amorphous solids~\cite{Edwards,Steinhardt1979,Steinhardt1981,Moshe}. Those studies did not, however, consider the issue of ``defects'' in amorphous solids in relation to non-affine displacements. 
From the point of view of non-affine deformations, it is evident that each and every atom in an amorphous solid plays, to some extent, the role of a topological defect, as we will show below.
The crucial difference with all previous studies, however, lies in the fact that the topological defects are not (and, in fact, they hardly can be) identified in the undeformed structure of the material, but rather in its displacement field under deformation.

We build a connection between non-affine deformations and topological defects here for the first time, by showing that the non-affine displacement field is indeed equivalent to the displacement field around a dislocation. In Fig.\ref{fig:liquids3}, we present a schematic Burgers circuit around a reference atom in an amorphous system (solid or liquid).  Any Burgers circuit around any atom, that is not a center of symmetry, in an amorphous solid leads to a non-zero Burgers vector $b_{i}$. 
This, therefore, implies that 
\begin{equation}
\oint_{L}du_{i}=\oint_{L}\frac{\partial u_i}{\partial x_{k}}dx_{k}=-b_{i}
\end{equation}
which can be recast in differential form using Stokes' theorem, as
\begin{equation}
\mathfrak{e}_{ilm}\partial_{l}\partial_{m}u_{k}=-\alpha_{ik}\,,
\end{equation}
where $\alpha_{ik}$ is the so-called dislocation density tensor, related to the Burgers vector via $db_{i}=\alpha_{mi}dA_{m}$, with $A_{m}$ being the axial vector orthogonal to the area element enclosed by the path $L$~\cite{Landau_elasticity,Tartaglia} and $dA_m$ the corresponding area element.

The above condition, in the language of elasticity theory, implies that the deformation is incompatible (see further below for more details about incompatibility). The possibility that non-affine deformations belong to the class of incompatible deformations has been suggested in Ref.~\cite{Zimmerman}. The above Fig.\ref{fig:liquids3} and the resulting analysis demonstrate that not only the non-affine deformations are incompatible, as expected, but also that every atom in an amorphous solid plays the role of a topological defect leading to a multi-valued displacement field. The non-affine displacement $u'_{i}$ is in every point of the material proportional to the Burgers vector, based on the construction of Fig.\ref{fig:liquids3}.

\section{The Frenkel theory of liquids}\label{sec3}
It is common wisdom to distinguish solids and fluids by the presence of propagating shear waves \cite{landau2013fluid}. In particular, it is often said that liquids have a vanishing low frequency shear modulus $G=0$ and therefore transverse phonons cannot propagate, and are replaced by a diffusive mode. This observation can be supported theoretically by stating that liquids are invariant under volume preserving diffeomorphisms, which indeed do protect the shear modulus from being finite \cite{Nicolis:2015sra,Dubovsky:2011sj} (see also \cite{Alberte:2015isw,Baggioli:2019abx} for a holographic verification of this statement). In other ways, the presence of a finite rigidity can be interpreted as the explicit breaking of such internal symmetry. Should we take this as the end of the story? Recent experiments have shown the existence of propagating shear waves at low frequency in confined liquids \cite{Noirez_2012} together with the associated solid-like elastic effects \cite{Kume2020,kume2020unexpected} (see also \cite{PhysRevLett.62.2616}). The theoretical discovery of the finite elasticity of confined liquids has been put forward using the framework of non-affine elasticity of structurally disordered condensed matter systems~\cite{Zaccone_PNAS_2020}. The key point is that confinement along one spatial direction effectively cuts off low-energy eigenmodes of the system which are responsible for the non-affine softening contribution to the shear modulus \cite{Zaccone_PNAS_2020,phillips2020universal}. The infrared cutoff due to confinement thus make the non-affine contribution effectively smaller, and leads to a finite low-frequency shear modulus in liquids with good wetting to the solid boundaries (the latter boundary condition is required to ensure the propagation of low-frequency phonons).

\subsection{Theoretical background}
A possible theoretical explanation for the vibrational excitations in liquids goes under the name of \textit{k-gap theory}
\cite{BAGGIOLI20201}. In this framework, the dynamics of transverse excitations is based on the coexistence of oscillatory solid-like vibrations together with the diffusive rearrangements of molecules. Defining the average time for these re-arrangements to be $\tau$, the dynamics of the shear modes is modified as follows:
\begin{equation}
    \omega^2\,+\,\frac{i\,\omega}{\tau}\,-\,v^2\,k^2\,=\,0\,.\label{kgap}
\end{equation}
One could immediately verify that whenever these local re-arrangements are ignored, i.e. taking the limit $\tau=\infty$, one recovers the standard propagating nature of shear sound waves. In the opposite limit, when $\omega \tau \ll 1$, the physics is described by a single hydrodynamic diffusive mode $\omega=-i\,v^2\,\tau\,k^2$.
\begin{figure}[ht]
    \centering
    \includegraphics[width=0.8\linewidth]{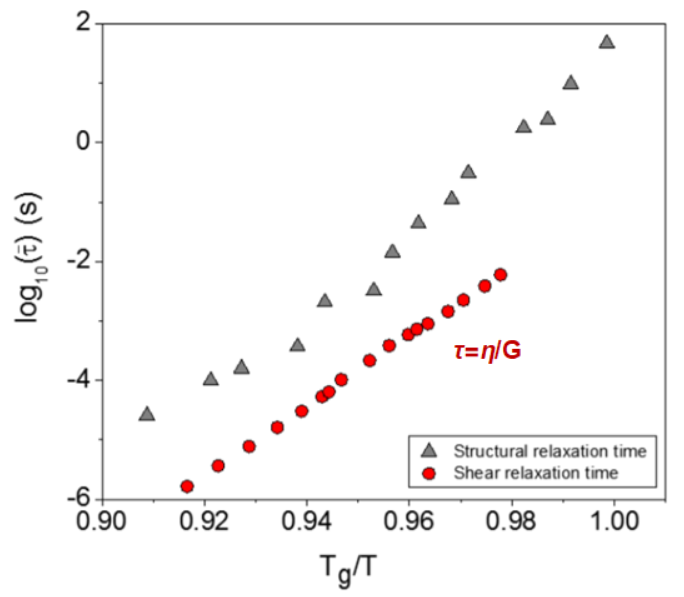}
    \caption{The clear discrepancy between the collective structural relaxation time and the single-particle Maxwell time $\tau=\eta/G$. Figure adapted from \cite{lancelotti2020does}.}
    \label{prova}
\end{figure}\\
Although this simple equation seems to work pretty well for several liquids \cite{PhysRevLett.92.065001,PhysRevLett.85.2514,PhysRevLett.84.6026,PhysRevLett.97.115001,Hosokawa_2015,doi:10.1063/1.5088141,PhysRevLett.125.125501,PhysRevLett.118.215502,PhysRevB.101.214312,doi:10.1063/1.5050708}, its derivation is purely phenomenological and based only on the physical intuition of Maxwell and Frenkel. In particular, the identification of the timescale $\tau$ is not obvious and this is mostly due to the absence of a theoretical and formal derivation of the Cattaneo-type equation \eqref{kgap} \footnote{See also \cite{ISRAEL1979341} for another phenomenological use of this equation in the context of linearized relativistic hydrodynamics.}. A (maybe too) simple derivation of this relaxation time exists and it can be obtained using the Maxwell linear viscoelastic theory. In this scenario, $\tau$ is identified as $\tau=\eta/G_{\infty}$ where $\eta$ is the shear viscosity and $G_{\infty}$ the instantaneous shear modulus -- the rigidity at infinite frequency. Unfortunately, this interpretation is highly debated and there is increasing contrary evidence from experiments \cite{lancelotti2020does} (see Fig.\ref{prova}) and from holographic results \cite{Andrade:2019zey,Baggioli:2018nnp,Baggioli:2018vfc} that contradict or question this interpretation.\\

Moreover, from a hydrodynamic point of view \'a la Landau, the introduction of a finite shear viscosity does not induce dynamics described by equation \eqref{kgap} but it rather provides solely a finite attenuation for the shear sound waves without any k-gap \cite{chaikin2000principles}. In particular, the relaxation scale $\tau$ appearing in Eq.\eqref{kgap} is the relaxation time of the collective hydrodynamic modes and not of the single liquid particles (atoms or molecules). A finite shear viscosity does not induce any relaxation time for the collective shear modes. On the contrary, the Frenkel time seems to attain to the dynamics of the single liquid molecule which is definitely not the character in discussion here since hydrodynamics refers only to the long wavelength collective dynamics. These points have been already discussed in the literature~ \cite{PhysRevLett.120.219601,doi:10.1021/acs.jpcb.8b01900,bryk2020comment}.\\
Recently, Ref.\cite{mokshin2021quasi} raised the problem that to quantify the ”k-gap” phenomenon
in liquids it is not sufficient to know only a single critical wave number $k_g$ determining the k-gap scale, because there is a crossover range of a finite width
in momentum space.\\

In this manuscript, we try to formulate a formal and consistent hydrodynamic and effective field theory description for shear waves in liquids giving a collective dynamics of the form \eqref{kgap}. As we will explain in the following, the only hydrodynamic consistent treatment producing k-gap dynamics for shear waves consists in the introduction of Goldstones phase relaxation, induced by the non-affine dynamics in liquids (and glasses).\\

Finally, let us emphasize that an equation like \eqref{kgap} correctly approximates the low-energy dynamics only in presence of a strong separation of scales for which the first two modes, the one appearing in \eqref{kgap}, are much ``lighter'' (in the sense of longer-living) with respect to the rest of the modes. This situation is nevertheless natural in the sense that sending to zero the damping of the second non-hydrodynamic mode in \eqref{kgap} produces a symmetry enhancement in the system.

\subsection{Physical picture}
Despite the derivation of the k-gap theory and Eq.\eqref{kgap} might be formally inaccurate and not consistent with the fundamental symmetries of liquids, the physical picture proposed by Frenkel \cite{frenkel} and then re-iterated by Trachenko and collaborators \cite{Trachenko_2015} is indeed correct, once the right mechanism of relaxation is identified. The phase relaxation rate $\Omega$, which is central in our theory, provides such a mechanism and relates microscopically to non-affine displacements in amorphous systems -- surprisingly, it has been totally ignored by previous works. $\Omega$ determines indeed the propagation length $\mathfrak{l}$ of the Goldstone bosons -- the collective phonon modes. Phase relaxation indicates that the Goldstone fields decay exponentially as (this is the physical content of the relaxed Josephson equation) :
\begin{equation}
    \Phi(t)\,\sim\,e^{-\,\Omega\,t}
\end{equation}
propagating therefore up to a length:
\begin{equation}
    \mathfrak{l}\,\equiv\,\frac{v}{\Omega}\,.
\end{equation}
Very importantly, and here it is where previous formulations fail, the relaxation time determining the Goldstones lifetime is not the momentum relaxation time and it is not an effect of the shear viscosity as in Maxwell's theory of viscoelasticity.

\begin{figure}[ht]
    \centering
    \includegraphics[width=\linewidth]{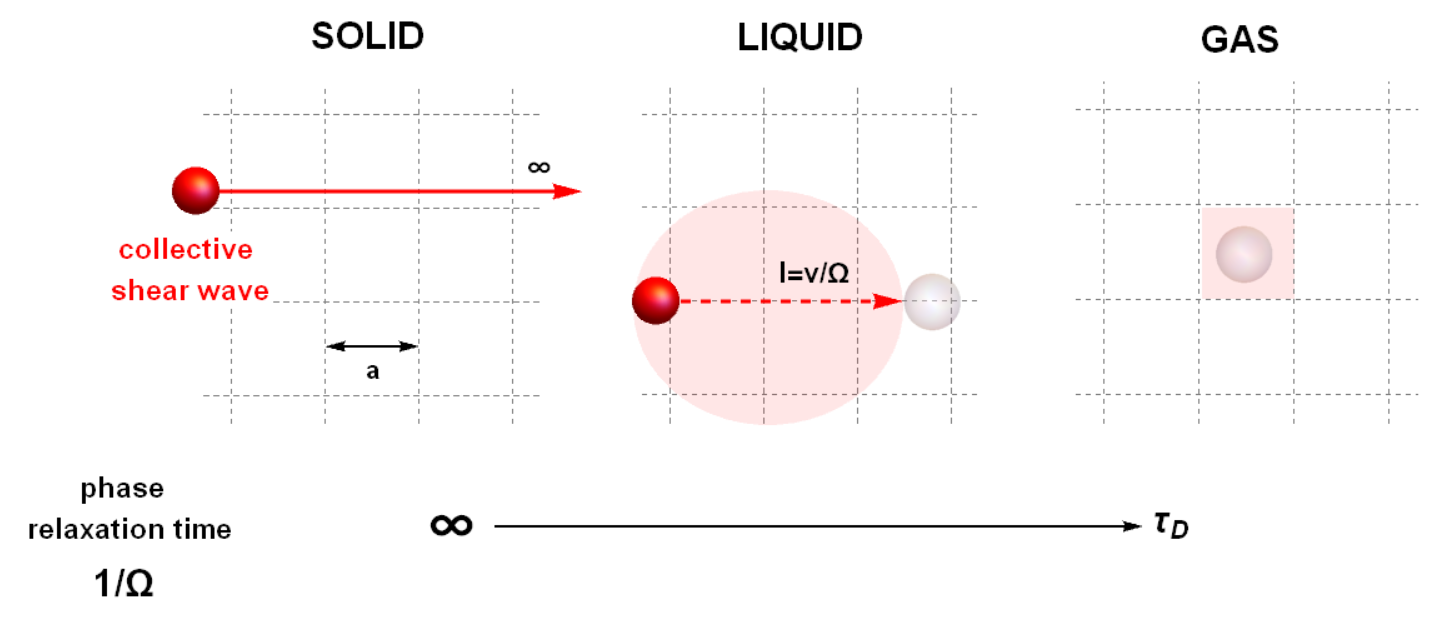}
    \caption{The correct physical picture corresponding to the k-gap theory. The lifetime, and consequently the propagation length, of the collective transverse phonons are determined by the phase relaxation time $1/\Omega$ which decreases increasing the temperature and the amount of non-affine displacements. Notice that within the k-gap theory the size $l$ is determined by the (inverse) of the k-gap scale $k_g$ in analogy with the propagation of electromagnetic waves in matter (see Section 5 of \cite{BAGGIOLI20201}).}
    \label{cartone}
\end{figure}

In ordered solids, non-affine deformations are practically absent, and the phase relaxation rate is zero, meaning that collective phonons propagate up to large distances and therefore low (theoretical zero) momentum. In liquids, after a time $1/\Omega$, the Goldstone modes are lost because the coherence of the phase is destroyed by the non-affine dynamics. One can therefore think of a liquid as a system in which the Goldstone dynamics in confined in clusters of size $\mathfrak{l}$. At lengths larger than that, the system loses completely the propagating shear waves and the associated rigidity. In the extreme scenario, when the relaxation time $1/\Omega$ reaches the Debye time $\tau_D$, then all transverse collective shear waves are lost. No collective mode can propagate whatsoever, since the minimum wave-length allowed is the Debye length, but at that point the propagation length is shorter than that. See Fig.\ref{cartone} for an illustration of this point.
\\

To conclude, there is an interesting analogy between this behaviour and the physics of superconductors, which has already been discussed in \cite{BEEKMAN20171}. Consider a type-II superconductor, above a certain critical magnetic field $B_{C1}$, it exhibits an intermediate phase of mixed ordinary and superconducting properties. In that phase, magnetic field lines start to penetrate into the superconductor and Abrikosov vortices, producing phase relaxation, start to proliferate in the medium. By increasing the magnetic field, the density of vortices increases, and with it the phase relaxation rate $\Omega$. The equivalent of the phonons propagation length is the magnetic field penetration depth, which has to be compared with the superconducting coherence length. The stiffness of the superconductor is highly affected by this process and above a second critical magnetic field $B_{C2}$, the superconductivity is completely destroyed. This last point would correspond to the energy at which the phonon propagating length reaches the Debye scale. The idea is to interpret the superfluid stiffness as the elastic rigidity and the phonons as the superfluid Goldstones, in perfect agreeement with P. W. Anderson's ideas on ``generalized rigidity''. A similar scenario, where vortices are created, are superfluids with flow \cite{RevModPhys.87.803}. It would be nice to make this comparison more precisely and exploit it to understand better the phase diagram of liquids.

\section{Hydrodynamics of systems with phase relaxation}\label{sec4}
Hydrodynamics (not to be confused with fluid-dynamics in the sense of Navier Stokes equations \cite{landau2013fluid}) is a universal low energy effective field theory built through a perturbative expansion in gradients. In particular, it describes the collective dynamics of a system at small frequencies and momenta, compared to the characteristic energy scale of the system. In a thermal system at equilibrium, the energy is given by the temperature $T$, and the hydrodynamic regime is confined to the window:
\begin{equation}
    \frac{\omega}{T}\ll\,1\,\,\,\,\text{and}\,\,\,\,\frac{k}{T}\,\ll\,1\,, \label{window}
\end{equation}
which we label as the \textit{hydrodynamic regime}.  In the equation above, and in the rest of the manuscript, we adopt Planck units $\hbar=k_B=c=1$. In a stricter sense, hydrodynamics governs the dynamics of conserved quantities and it is customary to define as \textit{hydrodynamics modes} those whose dispersion relation obeys:
\begin{equation}
    \lim_{k\,\rightarrow\,0}\,\omega(k)\,=\,0\,. \label{strict}
\end{equation}
As one can immediately see, the hydrodynamic regime is more general than this last definition. Sometimes, modes which do not obey \eqref{strict} but do lie within the region \eqref{window} are denoted as \textit{quasi-hydrodynamics}. This is for example the case for the instantaneous normal modes (INMs) typical of liquids \cite{Keyes,Stratt,Nitzan,murillo}.  An immediate consequence is that hydrodynamics, with the meaning explained above, can apply to the most disparate systems and not only to liquids. See for example the original work \cite{PhysRevA.6.2401}, whose title is quite eloquent in this context.\\

Here, we consider the hydrodynamic description of systems with spontaneously broken translations but with phase relaxation, which can be found for example in \cite{Delacretaz:2017zxd} (see also the Appendices of \cite{Ammon:2019apj} for more details). We follow the standard Kadanoff-Martin procedure \cite{KADANOFF1963419}. The starting point are the hydrodynamic variables, identified as the fields:
\begin{equation}
   \psi^A:\,\,  \epsilon(t,x)\,,\,\pi_\perp(t,x)\,,\,\pi_\parallel(t,x)\,,\,\lambda_\perp(t,x)\,,\,\lambda_\parallel(t,x)\,,
\end{equation}
the energy density, the momentum density parallel and transverse to the momentum; and the curl and divergence of the Goldstone
field $\phi_I(t,x)$ respectively ($\lambda_\perp= \nabla \times \phi$, $\lambda_\parallel= \nabla \cdot \phi$). The Goldstone fields are simply the linearized displacements and indeed the strain tensor can be defined as $\epsilon_{ij}=\partial_{(i}\phi_{j)}$. In this language, the free energy of an elastic crystal is just given by:
\begin{equation}
    f\,=\,\frac{1}{2}\partial_{(i}\phi_{a)}\,C^{ijab}\,\partial_{(j}\phi_{b)}\,+\,\dots
\end{equation}
with $C^{ijab}$ the standard elastic tensor.\\
The corresponding sources for the fields above are given by:
\begin{equation}
  s^B:\,\,   T(t,x)\,,\,v_\perp(t,x)\,,\,v_\parallel(t,x)\,,\,\Upsilon_\perp(t,x)\,,\,\Upsilon_\parallel(t,x)\,,
\end{equation}
where $T$ is the temperature, $v$ the velocity and $\Upsilon$ the sources of the Goldstone fields. The relations between the hydrodynamic variables $\psi^A$ and the sources $s^B$ are mediated through the matrix of thermodynamic susceptibilities $\chi_{AB}$:
\begin{equation}
    \psi^A\,=\,\chi^{AB}\,s_B\,.
\end{equation}
The full dynamics of the system is determined by the conservation of energy, the conservation of momentum and the Josephson relation for the Goldstone modes. In particular, the conservation of the stress energy tensor of the theory $T^{\mu\nu}$ (given by the invariance under time and spatial translations), reads:
\begin{equation}
    \partial_t\,\epsilon(t,x)\,+\,\partial^i\,T_{ti}(t,x)\,=\,0\,,\quad \partial_t\,\pi_i(t,x)\,+\,\partial^j\,T_{ij}(t,x)\,=\,0\label{eqzero}
\end{equation}
where $\epsilon \equiv T^{tt}$ and $\pi^i=T^{ti}$.\\
The constitutive relation for the stress tensor, at leading order in derivatives are given by:
\begin{align}
    T_{ti}(t,x) =& \chi_{\pi\pi} v_i(t,x) - \Bar{\kappa}_0 \,\partial_i T(t,x) \nonumber \\&- T(t,x) \,\gamma_2\, \partial_i \Upsilon_\parallel(t,{x}) + \mathcal{O}(\partial^2) \,,
\end{align}
and:
\begin{equation}
\begin{split}
      T_{ij}(t,{x}) &= \delta_{ij}\left[ p(t,{x}) - (K + G)\, \partial\cdot\phi(t,{x}) \right] \\ &\quad -2\,G\left[ \partial_{(i}\phi_{j)}(t,{x}) - \delta_{ij}\, \partial\cdot\phi(t,{x}) \right]\\ &\quad- \sigma_{ij}(t,{x}) + \mathcal{O}(\partial^2)\,,  
\end{split}
\end{equation}
with:
\begin{align}
    \sigma_{ij}(t,{x}) = & \eta \left(\partial_i v_j(t,{x}) + \partial_j v_i(t,{x}) - \delta_{ij} \partial_k v^k(t,{x})\right)\,\nonumber \\  & +\,\zeta\,\delta_{ij} \partial_k v^k(t,{x})\,.
\end{align}
Here we have made use of the following coefficients: $\Bar{\kappa}_0$ the thermal conductivity, $K,G$ the bulk and shear elastic moduli, $\zeta,\eta$ the bulk and shear viscosities, $p(t,x)$ the thermodynamic pressure and $\gamma_2$ a higher order dissipative coefficient. The stress tensor conservation has to be supplemented by the so-called Josephson relations for the Goldstones which are given by:
\begin{align}
    & \partial_t{\lambda}_\perp(t,{x}) - \partial\times \vec{v}(t,{x}) - \xi_\perp \partial_i\partial^i \Upsilon_\perp(t,{x}) = -\Omega_\perp \lambda_\perp(t,x);\label{eqdue}\\
    &\partial_t{\lambda}_\|(t,{x}) - \partial\cdot \vec{v}(t,{x}) - \gamma_2\,  \partial_j\partial^j T(t,{x}) - \xi_\| \, \partial_k\partial^k \Upsilon_\|(t,{x}) \nonumber \\ &= -\Omega_\parallel\,\lambda_\parallel(t,x)\,.\label{eqtre}
\end{align}
where $\xi_{\perp},\xi_\parallel$ determine the Goldstone diffusion constants and $\Omega_\perp,\Omega_\parallel$ the lifetime of the Goldstone modes. The latter constant are fundamental in our scenario since they are the phase relaxation rates of the Goldstones which, as we will see later, phenomenologically encode the explicit breaking of a specific two-form symmetry.\\

The Goldstones fields $\phi_i$ are defined via the commutation rule with the momentum density $\pi_j$:
\begin{equation}
    \left[\phi_i(x),\pi_j(y)\right]\,=\,i\,\delta(x-y)\left[\delta_{ij}+\partial_j \phi_i\right]
\end{equation}
which is simply the statement that translations shift the Goldstone fields \cite{Delacretaz:2017zxd}. Taking into account that the Hamiltonian contains a term:
\begin{equation}
    \mathcal{H}\,=\,\pi^i\,v_i\,+\,\dots
\end{equation}
where $v_i$ is the velocity vector, conjugate to the momentum density. Then, the time dynamics of the Goldstone modes, at leading order in the gradients, is given by:
\begin{equation}
    \dot{\phi}_i\,=\,i\,\left[\phi_i,\mathcal{H}\right]=v_i
\end{equation}
which corresponds to the Josephson relation for the Goldstones at leading order in derivatives. The name is taken in analogy with the superfluid case, in which the Hamiltonian contains a term:
\begin{equation}
    \mathcal{H}\,=\,\mu\,n\,+\,\dots
\end{equation}
where $\mu,n$ are respectively the chemical potential and the charge density. Then, in the superfluid case, the Josephson relation is simply $\dot{\phi}=-\mu$ which is equivalent to the famous Josephson effect \cite{beekman2020theory}.\\
Physically, the Josephson relation is just the dynamical equation for the Goldstone modes. Mathematically this can be seen either as the breaking of the higher-form conservation equation or equivalently as the introduction of an additional relaxational term in the Josepshon relation as:
\begin{equation}
    \left(\partial_t+\Omega\right) \phi_i + \dots =0
\end{equation}
Dynamically, this implies that the Goldstone mode is now exponentially relaxing as $\phi\sim e^{-\Omega t}$. Physically, this situation is usually encountered in presence of dynamical defects. A typical example is that of elastic defects such as dislocations and disclinations in solids. Another is the presence of vortices in superfluids. As we will extensively discuss, nonaffine dynamics can be a third, and more comprehensive, scenario in which this mechanism appears when structural order is destroyed.\\

To continue, equations \eqref{eqzero}, \eqref{eqdue} and \eqref{eqtre} determine the full dynamics of our system. The idea is then to expand all the hydrodynamic variables considering small fluctuations around their equilibrium values:
\begin{equation}
    \psi^A(t,x)\,=\,\psi^A_{eq}(t,x)\,+\,\delta \psi^A(t,x)\,+\,\mathcal{O}(\delta^2)
\end{equation}
and obtain the linearized equations for the fluctuation fields $\delta \psi^A$. Those equations, in Fourier space, take the matricial form:
\begin{equation}
    \mathcal{D}_{AB}(\omega,k)\,\delta\psi^A\,=\,0\,, \label{ma}
\end{equation}
where $\mathcal{D}_{AB}(\omega,k)$ is the usually called kinetic matrix. Because of the symmetries of the system, transverse and longitudinal fluctuations decouple and correspond to two different sets of equations of the type \eqref{ma}. The related hydrodynamic modes can be easily found by solving the eigenvalues problem:
\begin{equation}
    \det    \mathcal{D}_{AB}(\omega,k)\,=\,0\,.
\end{equation}
\begin{figure}[ht]
    \centering
    \includegraphics[width=0.45\linewidth]{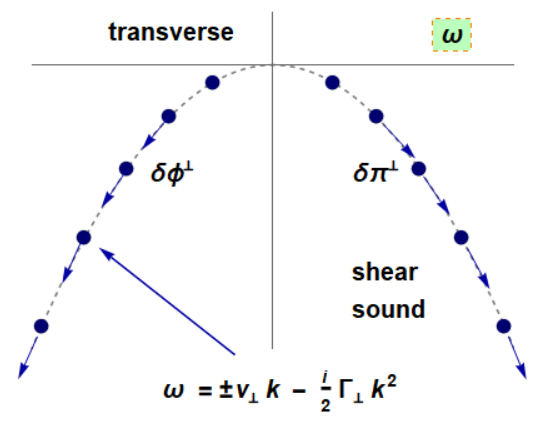}
    \qquad 
      \includegraphics[width=0.45\linewidth]{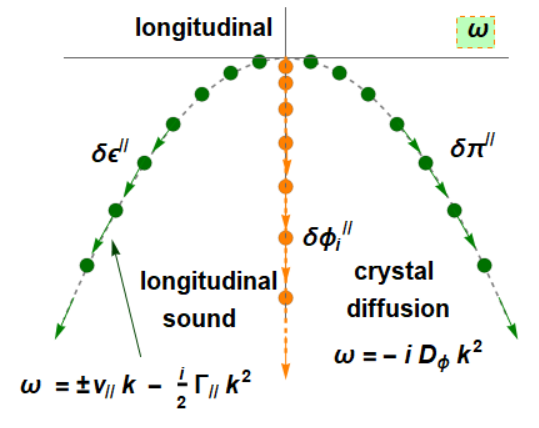}
      
      \vspace{0.5cm}
      
        \includegraphics[width=0.45\linewidth]{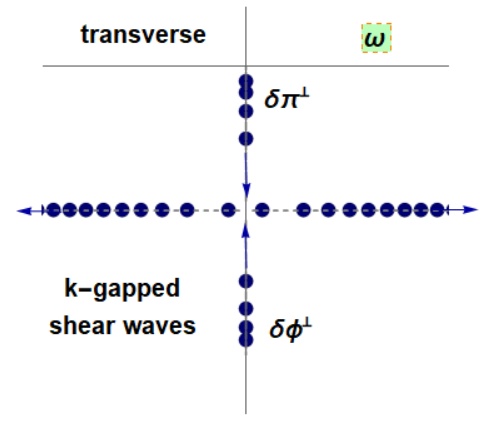}
        \qquad \includegraphics[width=0.45\linewidth]{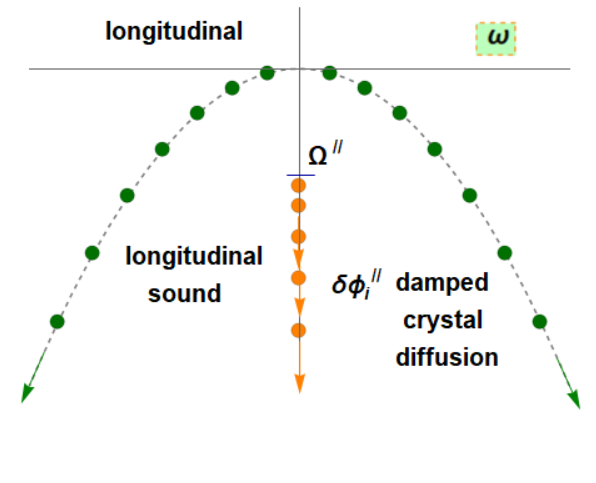}
    \caption{The dynamics of the hydrodynamics modes in the transverse and longitudinal spectrum. The bullets indicated the position of the modes in the complex plane $(\mathrm{Re}\,\omega,\,\mathrm{Im}\,\omega)$ and the arrows indicated their dynamics by increasing the momentum $k$. \textbf{Top panel: }The dynamics without phase relaxation. Notice how the longitudinal sound, contrarily from the transverse one, is not made of the corresponding longitudinal Goldstone field. \textbf{Bottom panel: } the dynamics with phase relaxation.}
    \label{fig:hydromodes}
\end{figure}\\
Ignoring the phase relaxation terms, and setting $\Omega_\perp=\Omega_\parallel=0$ , one recovers exactly the hydrodynamics spectrum of a perfect ordered crystal \cite{chaikin2000principles}. In particular, one finds the following modes:
\begin{itemize}
    \item tranverse sector: the shear sound waves with dispersion $\omega=\pm v_\perp k -\frac{i}{2}\Gamma_\perp k^2$.
    \item longitudinal sector: the longitudinal sound waves with dispersion $\omega=\pm v_\parallel k -\frac{i}{2}\Gamma_\parallel k^2$ and a crystal diffusive mode $\omega=-i D_\phi k^2$.
\end{itemize}

Here $v_\perp,v_\parallel$ are the sound speeds determined as usual by the elastic moduli $K,G$ and $\Gamma_\perp,\Gamma_\parallel$ the (Akhiezer) sound attenuation constants given in terms of the viscosities. Finally, $D_\phi$ is the crystal diffusion constant, which is determined by the dissipative parameter $\xi_\parallel$, and whose meaning is discussed in \cite{Baggioli:2020nay,Baggioli:2020haa}.\\

Now, once we introduce phase relaxation, the dispersion relation of the transverse shear waves gets modified as\footnote{Notice the higher order corrections $\mathcal{O}(k^4)$. This means that if the k-gap happens to be located at reasonably high momenta $k/T\sim 1$, then, the corresponding dynamics is not exactly described by the simple equation \eqref{kgap}. Still, the same type of dynamics will appear.}
\begin{align}
   & \omega_{\pm}\,=\,-\frac{i}{2}\,\Omega_\perp\,\nonumber  \pm\,\frac{1}{2\,\chi_{\pi\pi}}\,\mathcal{Q}\,,\\ &\mathcal{Q}\equiv \sqrt{k^2\,\chi_{\pi\pi}\,\left[4\,G\,-\,2\,\left(\xi_\perp-\eta\right)\Omega_\perp\right]\,-\,\chi_{\pi\pi}^2\,\Omega_\perp^2}\label{modes}
\end{align}
which neglects terms of order $\mathcal{O}(k^4)$ and which comes by imposing the determinant of the following matrix to vanish:
\begin{equation}
 \mathcal{D}^\perp_{AB}(\omega,k)\,=\,\left(
\begin{array}{cc}
 \frac{\eta\,  k^2}{\chi_{\pi\pi}}-i \,\omega  & -i \,G \,k \\
 -\frac{i\,k}{\chi_{\pi\pi}} & G \,k^2\, \xi_\perp -i \,\omega +\Omega_\perp  \\
\end{array}\,
\right)\,.
\end{equation}
Notice that neglecting all the dissipative terms ($\eta=\xi_\perp=\Omega_\perp=0$), one finds immediately $\omega^2=G/ \chi_{\pi\pi} \,k^2$, as expected for perfect non-dissipative solids.
\begin{figure}
    \centering
    \includegraphics[width=.4\linewidth]{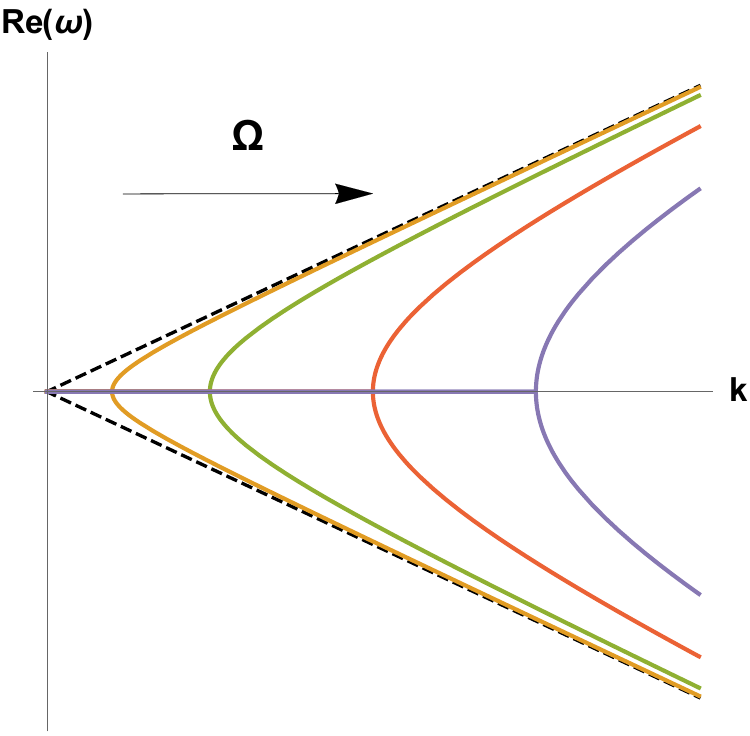}\quad \quad
      \includegraphics[width=.4\linewidth]{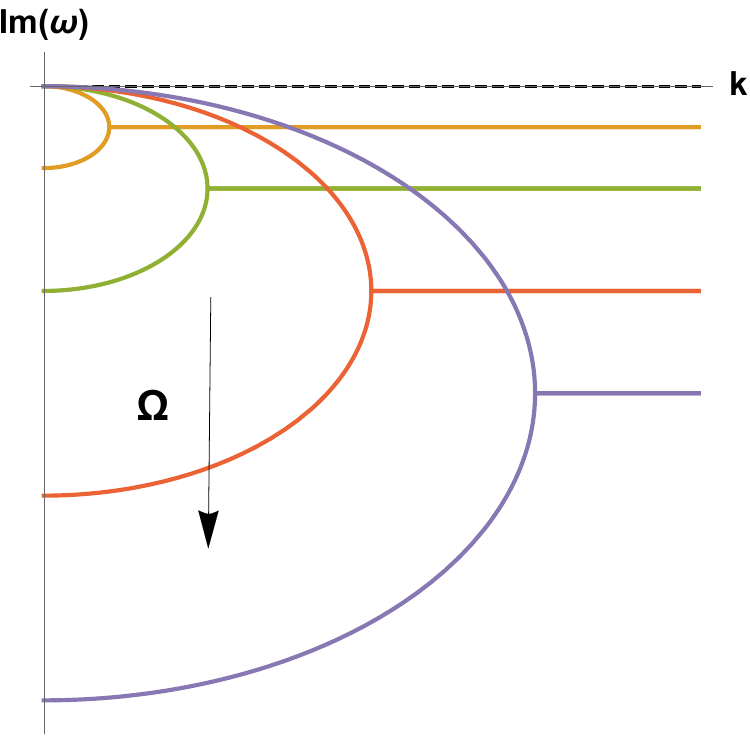}
    \caption{The dispersion relation of the collective shear waves in a system with phase relaxation Eq.\eqref{modes}. The dashed line is the result in absence of phase relaxation.}
    \label{fmodi}
\end{figure}\\
At low momentum, expanding \eqref{modes} for $k \ll 1$, we have two modes in the transverse sector:
\begin{equation}
    \omega_-=-i\, \Omega_\perp\,+\,i\,D\,k^2+\dots\,,\qquad \omega_+\,=\,-i\,D\,k^2\,+\,\dots
\end{equation}
where the diffusion constant is given by:
\begin{equation}
    D\,=\,\frac{2\,G\,+\,\left(\eta-\xi_\perp\right)\,\Omega_\perp}{2\,\chi_{\pi\pi}\,\Omega_\perp}\,.
\end{equation}
Instead, at large momentum, we have a pair of propagating transverse phonons:
\begin{equation}
    \omega\,=\,\pm\,D\,\Omega_\perp\,k\,-\,\frac{i}{2}\,\Omega_\perp\,+\,\dots
\end{equation}
The diffusive-to-propagating crossover point, which determines the k-gap scale is given by:
\begin{equation}
    k_g^2\,=\,\frac{\Omega_\perp}{D}\,.
\end{equation}
Following the same prescription, the hydrodynamics modes in the longitudinal sector can be found from the eigenvalues of the longitudinal kinetic matrix $\mathcal{D}^\parallel_{AB}(\omega,k)$:
\begin{equation}
    \left(
\begin{array}{ccc}
 \frac{\kappa_0\, k^2}{c_v}-i \,\omega  & i \,k &
   \gamma_2\, k^2\, T\, (G+K ) \\
 i \,\frac{dp}{d\epsilon}\, k & \frac{\eta \, k^2}{\text{$\chi $PP}}-i\,
   \omega  & -i\, k\, (G+K ) \\
 \frac{\gamma_2\, k^2}{c_v} & -\frac{i \,k}{\chi_{\pi\pi}} & k^2 \,\xi_\parallel\, (G+K )-i\, \omega +\Omega_\parallel  \\
\end{array}
\right)\label{bigm}
\end{equation}
where $G,K$ are the shear and bulk moduli, $c_v$ the specific heat at constant volume and for simplicity we have neglected the effects of a finite bulk viscosity by setting $\zeta=0$. The expressions for the corresponding modes are quite lengthy and therefore not shown explicitly.\\

Let us know analyze the properties following from Eq.\eqref{modes} and the eigenvalues of \eqref{bigm}. Starting from the transverse sector, the presence of phase relaxation ($\Omega_\perp \neq 0$) "kills" the transverse Goldstone modes and as a consequence the transverse sound at small momentum. This is shown explicitly in Fig.\ref{fmodi} for increasing phase relaxation rate. The dynamics is exactly of the k-gap type \cite{BAGGIOLI20201} with a well-defined diffusive-to-propagating crossover, whose location is determined by the phase relaxation rate $\Omega_\perp$.\\

\begin{figure}
    \centering
    \includegraphics[width=\linewidth]{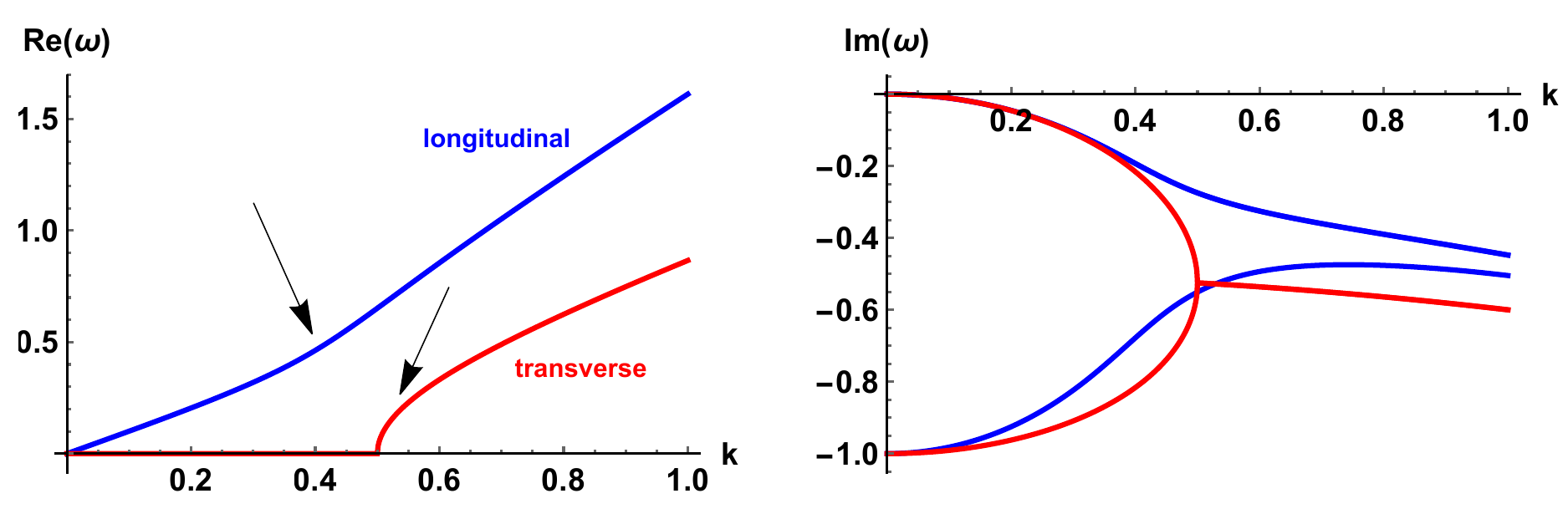}
    \caption{A comparison between the longitudinal modes (blue) and the transverse ones (red) at fixed phase relaxation rate $\Omega$. Here for simplicity we have assumed isotropy and therefore $\Omega_\perp=\Omega_\parallel$. The correlation between the transverse k-gap and the crossover to longitudinal fast sound is evident.}
    \label{tutto}
\end{figure}

\begin{figure}[t!]
    \centering
    \includegraphics[width=0.9\linewidth]{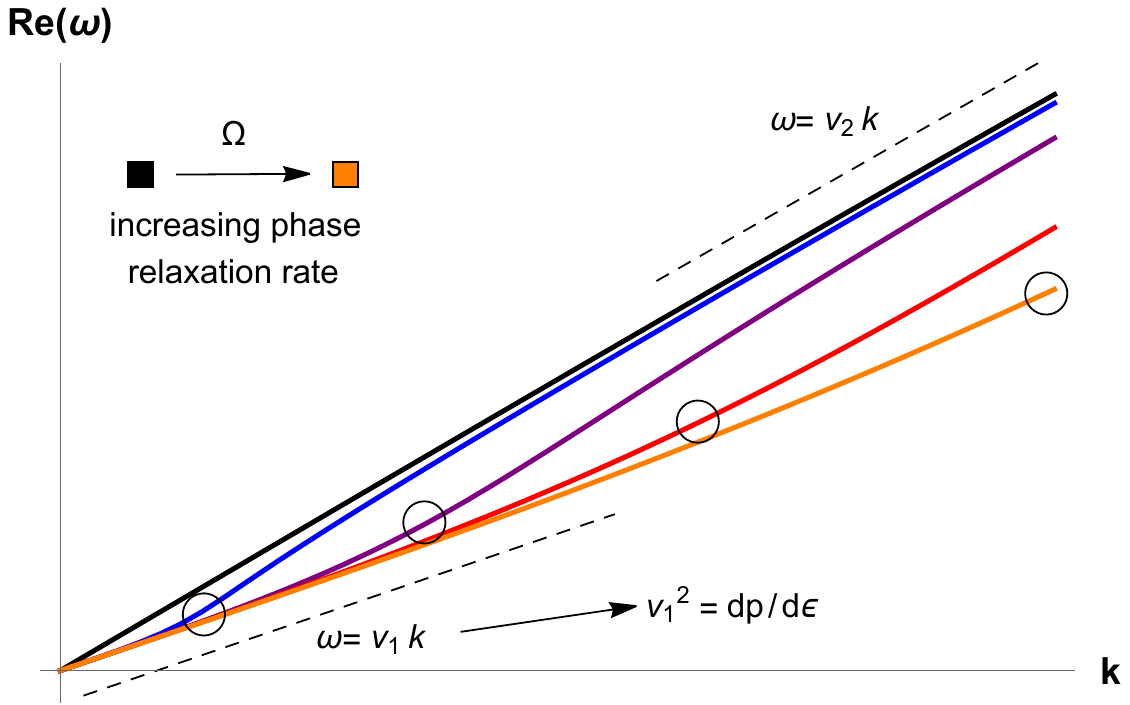}
    \caption{The dispersion relation of the longitudinal sound waves increasing the phase relaxation rate $\Omega$ and keeping all the other parameters fixed. The empty circles indicate the position of the crossover between slow and fast waves.}
    \label{fig:fast}
\end{figure}

\begin{figure}
    \centering
    \includegraphics[width=0.8\linewidth]{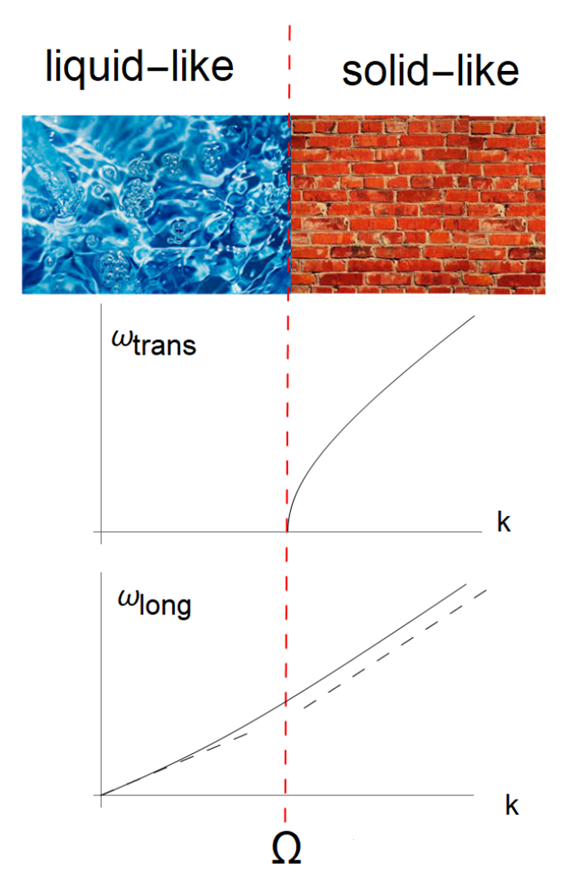}
    \caption{The effect of phase relaxation on the transverse and longitudinal dispersion relation of the sound modes. The transverse sound develops a gap; the longitudinal one a crossover between a slow regime and a fast one. The scale is controlled in both cases by the relaxation rate $\Omega$.}
    \label{sum}
\end{figure}

Conversely, in the longitudinal spectrum the sound modes are not "killed" by phase relaxation and despite their speed gets modified they still propagate at any value of momentum $k$. The reason for that is the following. As shown in Fig.\ref{fig:hydromodes}, the longitudinal sound comes from the mixing of energy fluctuations and longitudinal momentum fluctuations and it is independent of the fluctuations of longitudinal Goldstone field (which gets relaxed by the phase relaxation term). At low momentum, the longitudinal sector displays a pair of sound modes $\omega=\pm \sqrt{\frac{dp}{d\epsilon}}\,k$ and a damped mode $\omega=-i \Omega_\parallel$. At intermediate momenta, at a specific scale governed by the longitudinal phase relaxation rate $\Omega_\parallel$, there is a crossover to a second linearly dispersing regime with a faster speed. This crossover is evident from Fig.\ref{tutto}. Assuming isotropy, and therefore the parallel and transverse Goldstone diffusion constants and relaxation rates to be the same, this crossover is very close to the k-gap momentum in the transverse sector.
\\
The increase of the measured
speed of sound over its hydrodynamic value $\sqrt{\frac{dp}{d\epsilon}}$ is often labelled “fast sound” or “positive sound dispersion” (PSD). It has been observed in experiments and simulations \cite{PhysRevA.34.602} and discussed at length in the literature \cite{Trachenko_2015}. Here, we show that the origin of this phenomenon is again to be found in the phase relaxation mechanism induced by the liquid non-affine dynamics.\\
In particular, looking at Fig.\ref{fig:fast}, we can notice how the presence of a finite relaxation rate $\Omega$ affects the dispersion relation of longitudinal sound. At zero relaxation rate, the speed of longitudinal sound is given by simple expression:
\begin{equation}
    \omega=\pm v_2 \,k\,,\qquad v_2^2\,=\,\frac{dp}{d\epsilon}\,+\,\frac{G+K}{\chi_{\pi\pi}}\,>\,\frac{dp}{d\epsilon}\,.
\end{equation}
\\
At finite relaxation rate, a second linear regime appears at small momentum, with dispersion relation:
\begin{equation}
    \omega=\pm v_1 \,k\,,\qquad v_1^2\,=\,\frac{dp}{d\epsilon}\,,
\end{equation}
where $v_1$ is exactly the speed of sound in a fluid with no elastic components. Increasing the phase relaxation rate, this "liquid regime" becomes wider and wider growing in the same way as the gap in the transverse sector. The reason why the longitudinal sound wave is not gapped is that the full bulk modulus $K_{tot}\equiv -V dp/dV$ is finite in liquids and therefore longitudinal sound waves propagate even in liquids (differently to transverse ones). As such, the crossover between "liquid"-to-"solid" behaviour is of different nature depending of the sector we are looking at. A summary of the physical picture is provided in Fig.\ref{sum}.

This is a further confirmation of the validity of our framework which explains from first principles the low energy dynamics of liquids both in the transverse and longitudinal sector without relying on magical interpolations \'a la Maxwell/Frenkel. It is interesting to observe that the same PSD phenomenon has been predicted in the framework of Hydro+ \cite{Stephanov:2017ghc} (see Fig.2 therein) considering the effects of hydrodynamics fluctuations and the presence of a slowly relaxing additional modes. The similarities with our picture are striking and definitely deserve more attention. Indeed, also in such formalism the crossover scale is given by the lifetime of the slowly relaxing additional modes $\Gamma_\pi$, which in our language is exactly $\Omega$, once the slowly relaxing mode is identified with the Goldstone mode.\\
Notice that in order to destroy the longitudinal sound, in the same way the phase relaxation rate does with the transverse one, we would need to introduce the explicit breaking of translations or energy conservation. This can be easily understood from the fact that to relax longitudinal sound we need to relax either longitudinal momentum fluctuations or energy fluctuations.\\

Before moving to the next section, let us make few remarks.\\
(I) From Eq.\eqref{modes}, the asymptotic (instantaneous) speed of the shear waves is given by:
\begin{equation}
    v^2\,\equiv\,\frac{G_\infty}{\chi_{\pi\pi}}\,=\,\frac{G}{\chi_{\pi\pi}}\,+\,\frac{\left(\eta-\xi_\perp\right)\,\Omega_\perp}{2\,\chi_{\pi\pi}}\,.
\end{equation}
where $G$ indicates the static (zero frequency) shear modulus, sometimes labelled as $G_0$.\\
In the regime $\eta> \xi_\perp$ \footnote{It would be interesting to see if there is any fundamental physical constraint behind this inequality.}, this implies that:
\begin{equation}
    G_\infty\,=\,G\,+\,\frac{\left(\eta-\xi_\perp\right)\,\Omega_\perp}{2}\,>\,G\,,
\end{equation}
This equation can also be used to express the zero-frequency shear modulus as:
\begin{equation}
    \,G\,\,=G_\infty-\,\frac{\left(\eta-\xi_\perp\right)\,\Omega_\perp}{2}\,<\,G_\infty\,,
\end{equation}
where the negative (softening) term is related to non-affine displacements through the phase relaxation $\Omega_\perp$. The above form of $G$ is consistent with the microscopic result of non-affine elasticity theory, cfr. Eq. (6) by Zaccone and Scossa-Romano, presented above in Sec. 2.2. This consistency check again demonstrates that the microscopic origin of phase relaxation in fluids and glasses coincides with the microscopic non-affine displacement field and its topological properties.

In a perfect ordered crystal with no defects, one would have $G_\infty=G$, as expected. The presence of defects, and non-affine displacements in non-centrosymmetric environments, diminish the value of the static shear modulus which could eventually vanish at marginal stability (e.g. glass transition, jamming transition and alike):
\begin{equation}
    G\,=\,0\,.
\end{equation}
It is indeed a well known fact that in liquids $G=0$ but $G_\infty \neq 0$ and is large~\cite{MOUNTAIN1977225}.
In our language, that implies that phase relaxation is too strong and the Goldstones decay too fast to sustain any rigidity in the material.\\

(II) The presence of a phase relaxation term modifies the shear viscosity of the system as \cite{Delacretaz:2017zxd}:
\begin{equation}
    \Tilde{\eta}\,\equiv\,\lim_{\substack{\omega \rightarrow 0\\k \rightarrow 0}}\,\mathrm{Re}\,\frac{i\,\omega}{k^2}\,\mathcal{G}^{R}_{\pi_\perp \pi_\perp}(\omega,k)\,=\,\frac{G}{\Omega_\perp}\,+\,\eta \label{comp}
\end{equation}
This is interesting, because it could be compared with the phenomenological Maxwell-Frenkel interpolation \cite{maxwell,frenkel}:
\begin{equation}
    \frac{1}{\Tilde{\eta}}\,=\,\frac{1}{\eta}\,\left(1\,+\,\tau\,\frac{d}{dt}\right)
\end{equation}
which implies:
\begin{equation}
    \Tilde{\eta}\,\approx\,\eta\,+\,i\,\omega\,\tau\,\eta\,+\,\dots \qquad \text{for}\qquad \omega \tau \ll 1\,.
\end{equation}
The two approaches display a substantial difference. More precisely, the inclusion of phase relaxation modifies the shear viscosity even at zero frequency while Maxwell interpolation only at finite (and actually large $\omega \tau \sim 1$) frequency.\\
\section{A perspective from symmetries}\label{sec5}
\subsection{A dual formulation of elasticity theory}\label{subsub}
In order to connect robustly the presence of phase relaxation with the symmetries of the system, it is constructive to consider a dual formulation of the theory of elasticity based on a two-form global symmetry
\cite{Grozdanov:2018ewh} (see also \cite{BEEKMAN20171} for similar previous ideas). The main idea is to consider the Goldstone degrees of freedom in a solid as a set of massless and shift symmetry scalar fields $\phi^I$. At leading order, the action for the solids is given by:
\begin{equation}\label{dual action a}
    S\,=\,-\,\int d^3x\,{C^{\mu\nu}}_{IJ}\,\partial_\mu \phi_I \partial_\nu \phi_J
\end{equation}
implying that the conjugate momentum is given by:
\begin{equation}
    P_I^\mu\,=\,-\,{C^{\mu\nu}}_{IJ}\,\partial_\nu \phi_J\,.\label{w1}
\end{equation}
The conservation of momentum is then simply:
\begin{equation}
  \partial_\mu\,P^\mu_I\,\equiv\,  d* P_I\,=\,0\,.
\end{equation}
Nevertheless, there is another more subtle topological symmetry, which implies the conservation of the two-form current:
\begin{equation}
   \partial_\mu\,J^{\mu\nu}_I\,=\,0\qquad \text{with}\qquad J^{\mu\nu}_I\,=\,\mathfrak{e}^{\mu\nu\rho}\partial_\rho \phi_I\,, \label{duecons}
\end{equation}
with $\mathfrak{e}^{ijk}$ the anti-symmetric Levi-Civita tensor.\\
This conservation is equivalent to the fact that the scalars (displacements) $\phi^I$ are single-valued functions:
\begin{equation}
    \partial_{[\mu}\partial_{\nu]}\phi_I\,=\,0\,.
\end{equation}
The associated conserved charges:
\begin{equation}
    \mathcal{Q}_I\,=\,\int_{S^1}\,*\,J_I
\end{equation}
are the number of lattice sites in the elastic medium, counted across the surface of a circle ${S^1}$.\\
In this language, the conservation of momentum is the equivalent of the conservation of magnetic flux lines in electromagnetism and the conservation of the two-form current is simply the conservation of electric flux lines, the topological Bianchi identity.\\
Following \cite{Grozdanov:2018fic}, it is easy to show that the interplay of the two-form dynamics with the transverse momentum produces the presence of propagating transverse shear waves as expected in solids. For simplicity, let us show it in two spatial dimensions. By considering the fluctuations of the transverse momentum operator as:
\begin{equation}
    P_\perp(t,x)\,=\,T_{\mu y}(t,x)\,dx^\mu
\end{equation}
and using Eqs.\eqref{w1},\eqref{duecons}, it follows that:
\begin{equation}
    J_\perp(t,x)\,=\,\mathfrak{e}_{\mu\nu\lambda}\,\left(C^{-1}\cdot T\right)^{\lambda y}\,dx^\mu \wedge dx^\nu\,.
\end{equation}
The conservation of the momentum and of the two form $J$ implies then:
\begin{equation}
\begin{cases}
    &\partial_t T^{ty}\,+\,\partial_x T^{xy}\,=\,0\\
    & \partial_t\,\left(C^{-1}\cdot T\right)^y_x\,-\,\partial_x\,\left(C^{-1}\cdot T\right)^y_t\,=\,0
\end{cases}
\end{equation}
which gives rise to a propagating transverse wave -- the shear sound.
\subsection{Phase relaxation as the explicit breaking of a two-form symmetry}
We are now in the position to study how the introduction of phase relaxation modifies the conservation equation \eqref{duecons}.\\
Let us start from the phenomenological hydrodynamic treatment, which leads to the deformed Josephson relation written in the relaxation time approximation (RTA):
\begin{equation}
    \partial_t\,\lambda_\perp\,=\,-\,\Omega_\perp\,\lambda_\perp\,+\,\dots \label{jj}
\end{equation}
in terms of:
\begin{equation}
    \lambda_\perp^k\,=\,\left(\nabla \times \phi\right)^k\,=\,\mathfrak{e}^{klm}\,\partial_l\,\phi_m\,.
\end{equation}
Taking into account that $J^{\mu\nu}_I= \mathfrak{e}^{\mu\nu\rho}\partial_\rho \phi_I$, we find the equivalence:
\begin{equation}
    J^{\mu I}_{I}\,=\,\left(\nabla \times \phi\right)^\mu\,=\,\lambda_\perp^\mu\,.
\end{equation}
Now, in systems with only broken spatial translations (i.e. solids), the Goldstone fields have finite components only in the spatial directions. In particular, at equilibrium we can take $\langle \phi^I \rangle = x^I$, with the index $I$ running only on the spatial directions. Therefore, the only non-zero component of $\nabla \times \phi$ is the temporal one, implying $J^{tI}_I\neq 0$ only.\\
To conclude, we can therefore write the modified Josephson relation as:
\begin{equation}
    \partial_t \lambda_\perp\,=\,\partial_t\,J^{tI}_{I}\,=\,-\,\Omega\,J^{tI}_I
\end{equation}
where for simplicity we have considered only the transverse part. By taking carefully into account all the components and making the expressions covariant, one finally obtains:
\begin{equation}
    \partial_\mu\,J^{\mu\nu}_I\,=\,-\,\Omega\,J^{t\nu}_I\,.
\end{equation}
The latter implies that phase relaxation is equivalent to the explicit breaking of the two-form symmetry, as already advertised in \cite{Grozdanov:2018fic}.\\
Notice that the r.h.s. of the broken Ward identity for the two-form $J^{\mu\nu}$ assumes a very specific form. This specific form goes under the name of \textit{relaxation time approximation}, and it follows from the same assumptions made in Eq. \eqref{jj} at the level of the Josephson relation. Needless to say that this is not the most general symmetry breaking pattern one can imagine. In particular, this approximation relies on the fact that (I) the relaxation rate $\Omega$ is small and (II) that there is a well-defined separation of scales allowing us to define a single relaxation time. As we will see, at the level of the most general effective field theory description, it is easy to generalize these statements.
\subsection{Two-form symmetry breaking and compatibility equation}

The description of solids, the \textit{theory of elasticity}, is based on the displacements:
\begin{equation}
    u_i\,\equiv\,x'_i\,-\,x_i\,,
\end{equation}
which measure the geometrical deformations around equilibrium $x_i$, and by the corresponding strain tensor:
\begin{equation}
    \epsilon_{ij}\equiv\,\partial_{(i}u_{j)}\,.
\end{equation}
In its continuum formulation, it is fundamental that the various volume elements composing the rigid body are connected to each other without any gaps nor overlaps and that this condition is preserved during any mechanical deformations~\cite{Zimmerman}. This requirement takes the name of \textit{compatibility} and was first derived by
Barre de Saint-Venant (1864) \cite{M.1906}, and later rigorously proven by Beltrami in 1886 \cite{beltrami1886sull}.
The compatibility conditions correspond to the requirement of having a single-valued displacement field~\cite{Bassani,Zimmerman}. Whenever the strain can be assumed to be infinitesimal, these conditions are equivalent to the fact that the displacements can be obtained by integrating the strains and they can be expressed as
\begin{equation}
    \oint_{L}\,du_i\,=\,0\,.
\end{equation}
The integral of the displacement fields around a close loop $L$ must vanish.\\
In full generality, the displacement field can be written as:
\begin{equation}
    du_i\,=\,\left(\epsilon_{ij}+\omega_{ij}\right)\,dx_j
\end{equation}
where the first term -- the small strain tensor -- is the symmetric part and the second the anti-symmetric one. Then, after some algebraic manipulations, one can prove that:
\begin{equation}
    \oint\,\omega_{ij}\,dx_j\,=\,-\,\int x_l\,\omega_{ij,l}\,dx_j
\end{equation}
and moreover, using standard tensorial identities, that:
\begin{equation}
    \omega_{ij,l}\,=\,\mathfrak{e}_{mil}\,\mathfrak{e}_{mpq}\,\epsilon_{pj,q}
\end{equation}
where $\mathfrak{e}$ is the Levi-Civita tensor. All in all, we can now write:
\begin{equation}
    \oint du_i\,=\,\oint\,\left(\epsilon_{ij}\,-\,x_l\,\mathfrak{e}_{mil}\,\mathfrak{e}_{mpq}\,\epsilon_{pj,q}\right)\,dx_j
\end{equation}
which, using Stokes theorem $\oint F \cdot dx\,=\,\int \int_S\,n\,\cdot\,\nabla \times F\,dS$, becomes:
\begin{equation}
    \oint du_i\,=\,-\,\int\int_S\,n_r\,\mathfrak{e}_{mil}\,\left(\mathfrak{e}_{rsj}\,\mathfrak{e}_{mpq}\,\epsilon_{pj,qs}\right)\,x_l\,dS\,.
\end{equation}
The term inside the bracket is then the curl of the curl of the strain tensor:
\begin{equation}
    \mathfrak{e}_{rsj}\,\mathfrak{e}_{mpq}\,\epsilon_{pj,qs}\,\equiv\,\nabla \times \nabla\times \epsilon\,.
\end{equation}
Therefore, we end up discovering that the compatibility condition is equivalent to the requirement
\begin{equation}
    \nabla \times \nabla\,\times \epsilon\,=\,0\,,
\end{equation}
where $\epsilon_{ij}$ is the infinitesimal strain tensor.\\
Using the map in terms of the scalar fields $\phi^I$, the displacements are basically represented by the fluctuations of such scalar fields around their equilibrium position $\phi^I_{eq}=x^I$. This means that the curl of the strain tensor can be written as:
\begin{equation}
    \nabla \times \epsilon\,=\,\mathfrak{e}^{ijk}\partial_i\,\epsilon_{mj}\,=\,\mathfrak{e}^{ijk}\partial_i\,\partial_k\,\phi_m\,.
\end{equation}
The curl of the strain tensor being zero is then the same of the single-valued conditions
for the displacement fields:
\begin{equation}
    \left[\partial_i,\partial_j\right]\,\phi_k\,=\,0\,.
\end{equation}
Finally, we have already seen that this corresponds to the conservation of the dual two-form $\mathcal{J}$ as:
\begin{equation}
    \partial_\mu J^{\mu\nu}_I\,=\,\partial_\mu \,\mathfrak{e}^{\mu\nu\rho}\,\partial_\rho\,\phi_I\,=\,0
\end{equation}
and the presence of a finite Burgers vector:
\begin{equation}
    \oint du_i\,=\,-\,b_i\,.
\end{equation}
In short summary, we have revealed a deep and fundamental net of dualities which can be summarized as:

\begin{center}
{\centering\resizebox{9cm}{7cm}{%
\centering
\begin{tikzpicture}
\centering
  \path[mindmap,concept color=black,text=white]
    node[concept] {COMPATIBILITY}
    [clockwise from=45]
    child[concept color=green!50!black] {
      node[concept] {Zero Burgers vector}
      [clockwise from=45]
    }  
    child[concept color=blue] {
      node[concept] {Two-form symmetry}
    }
    child[concept color=red] { node[concept] {?} }
     child[concept color=red] { node[concept] {?} }
      child[concept color=purple] { node[concept] {No phase relaxation} }
    child[concept color=orange] { node[concept] {Single-valued displacements} };
\end{tikzpicture}}}\end{center}
In particular, the higher-form symmetry, imposing the conservation of the dual two-form $\partial_\mu J^{\mu\nu}_I=0$, is equivalent to the existence of single-valued displacements and the vanishing of the Burgers vector and it implies the vanishing of the phase relaxation rate $\Omega$. As a consequence, its explicit breaking:
\begin{equation}
    \partial_\mu\,J^{\mu\nu}_I\,\neq\,0
\end{equation}
corresponds to having topological defects in the system, which directly cause phase relaxation in the Goldstones dynamics.\\
The missing links, shown above as a question mark in the red bubbles, is what this breaking has to do with the condition of \textit{affine} displacements, and how to construct a formal effective field theory describing it. These are the topics of the next sections.\\

Finally, let us mention that there are more mathematical connections which we did not intend to explore. In particular, the previous story can be further connected with the mathematical property of \textit{homotopy} \cite{RevModPhys.51.591} and \textit{torsion} and the definition of Einstein-Cartan gravitational theories~\cite{Tartaglia}. For a complete discussion on these points see \cite{kleinert1989gauge}.

\section{From non-affine displacements to phase relaxation}\label{sec6}
In this section, we connect together all the properties discussed so far and in particular we draw a direct connection between the presence of non-affine displacements and the macroscopic hydrodynamic phase relaxation of the Goldstone modes, which is one of the main results of our work. Let us start by considering a generic mechanical deformation whose associated displacement field can be written as \cite{PhysRevE.72.066619}:
\begin{equation}
    u_i(\textbf{x})\,=\,\underbrace{\gamma_{ij}\,x^j}_{\text{affine}}\,+\,\underbrace{u'_i(\textbf{x})}_{\text{non-affine}}\,,
\end{equation}
where $\gamma_{ij}$ is a a constant tensor determining the affine component of the deformation. The second non-affine term $u'$ does not obey any specific requirements.\\
The first quantity we want to compute is the circulation of the displacement field, which defines the Burgers vector:
\begin{equation}
    \oint_L\,du_i\,=\,-\,b_i\,.
\end{equation}
The l.h.s. can be written using the tensor field $F_{ij}\equiv \frac{\partial u_i}{\partial x^j}$ as:
\begin{equation}
    \oint_L\,F\,\cdot\,ds\,=\,\oint_L\,\frac{\partial u_i}{\partial x_j}\,\cdot\,dx^j\,,
\end{equation}
where the tensor $F$ is simply:
\begin{equation}
    F_{ij}\,=\,\gamma_{ij}\,+\,\frac{\partial u'_i}{\partial x^j}\,.
\end{equation}
The first term in the r.h.s. -- the affine contribution -- is a constant term and it therefore defines a conservative field $\oint_L\,\gamma\,\cdot\,ds\,=\,0$. This also means that the associated strain tensor $\gamma_{ij}$ is irrotational and the corresponding deformation compatible. Consequently, the Burgers vector $b$ is solely determined by the non-affine contribution of the displacement as:
\begin{equation}
    \oint_L\,du'_i\,=\,-\,b_i\,. \label{equno}
\end{equation}
Now, let us define the non-affine part of the tensor $F_{ij}$ as $\mathcal{N}_{ij}\equiv \frac{\partial u'_i}{\partial x^j}$. Using Stokes theorem, we can re-write the l.h.s. of Eq.\eqref{equno} as:
\begin{equation}
    \int\int_S\,\nabla \times \mathcal{N}\,\cdot\,\hat{n}\,dS
\end{equation}
where $S$ is the surface enclosed by the loop $L$ and $\hat{n}$ the normalized vector perpendicular to such surface. Notice that this term is not zero since $\mathcal{N}$ is not irrotational -- it is indeed a non-compatible deformation -- and in index notations it reads:
\begin{equation}
    \int \int_S\,\mathfrak{e}^{abj}\,\partial_b\,\mathcal{N}_{ij}\,n_a\,dS\,.
\end{equation}
We can now assume that this integral is non-zero for any surface $S$ and vector $n$. This implies that:
\begin{equation}
    \mathfrak{e}^{abj}\,\partial_b\,\partial_j\,u'_i\,\neq\,0
\end{equation}
which is simply the statement that the Nye tensor \cite{NYE1953153}, measuring the density of elastic defects (non-affinity), is finite:
\begin{equation}
    \mathfrak{e}^{abj}\,\partial_b\,\partial_j\,u'_i\,\equiv\,-\,\alpha^a_{i}\,\neq\,0\,.
\end{equation}
Using the definition of the two form $J^{\mu\nu}$, this last expression can be re-written as:
\begin{equation}
    \,\alpha^a_{i}\,=\,\partial_\mu J^{\mu a}_i\,=\,-\,\Omega\,J^{ta}_i\,\neq\,0
\end{equation}
which indicates the presence of phase relaxation in the system.\\
Following this logic flow, we have explicitly demonstrated that \underline{non-affinity is equivalent to the} \underline{presence of a macroscopic Goldstone phase relaxation}
which closes our loop and connect directly our microscopic starting point with its macroscopic collective effect.
\section{Non-affinity, phase relaxation and higher-form symmetry breaking}\label{secnew}
Before moving to the next Sections, let us review our main claims so far along with their meaning. In the previous parts of this work, we have argued that:
\begin{enumerate}
    \item The appearance of a macroscopic phase relaxation $\Omega \neq 0$ is equivalent to the non-conservation of the two-form current $\partial_\mu J^{\mu\nu}_I\neq 0$.
    \item The presence of a finite Burgers vector $b_i\neq 0$ is equivalent to the non-conservation of the two-form current $\partial_\mu J^{\mu\nu}_I\neq 0$ and therefore via point (1) to a finite phase relaxation $\Omega$.
    \item Microscopically, the fundamental reason behind all of this is that the displacement field has a finite non-affine part, or equivalently is a multi-valued field.
\end{enumerate}
Let us briefly give a very explicit and concrete derivation of these main statements.\\

\textbf{Point (1)} -- Let us start by defining the charges $q^j_I$ as:
\begin{equation}
    q^j_i \equiv \int d^2x\,J^{0j}_i
\end{equation}
where $0$ stands for the time coordinate and $J^{\mu\nu}_i$ is the two-form defined in the previous Sections. For convenience, let us split the charges into:
\begin{align}
   & q_\perp\equiv \int d^2 x \,\delta^{ij} \,J^{0j}_i\,,\\
   &q_\parallel \equiv \frac{1}{2}\int d^2 x\,\epsilon^{ij}\,J^{0j}_i\,.
\end{align}
As in the more familiar $U(1)$ case, the higher-form symmetry implies the conservation of these two charges, $\dot{q}_\perp=\dot{q}_\parallel=0$.\\
Using the notations introduced above, we can verify that:
\begin{align}
   & \delta^{ij} \,J^{0j}_i= \epsilon^{ij} \partial_j u_i=\nabla \times u =\lambda_\perp\,,\\
   &\epsilon^{ij} \,J^{0j}_i=\nabla \cdot u = \lambda_\parallel\,.
\end{align}
Then, we have:
\begin{equation}
    \dot{q}_\perp=\int d^2x \dot{\lambda}_\perp\,,\qquad \dot{q}_\parallel=\int d^2x \dot{\lambda}_\parallel\,.
\end{equation}
Recalling that the presence of a macroscopic phase relaxation implies $\dot{\lambda}_\perp,\dot{\lambda}_\parallel \neq 0$, one immediately derives that the latter is equivalent to the non-conservation of the charges ${q}_\perp,q_\parallel$ and therefore to the breaking of the higher-form symmetry.\\

\textbf{Point (2)} -- The second step is realizing that:
\begin{equation}
    J_i = * \,du_i
\end{equation}
where $*$ is the standard 3-dimensional Hodge dual. This implies that:
\begin{equation}
    b_i=\oint_C du_i=\oint_C * J_i=\int_\Sigma d * J_i
\end{equation}
where the loop $C$ is the boundary of the surface $\Sigma$. Finally, we notice that:
\begin{equation}
    d * J_i=0 \,\, \longrightarrow \,\,\partial_\mu J^{\mu\nu}_i=0\,.
\end{equation}
This last equality implies immediately that a non zero Burger vectors $b_i \neq 0$ is linked 1-to-1 with the non-conservation of the two-form current $J^{\mu\nu}_i$.\\

\textbf{Point (3)} -- Let us stress that by $u_i$ being non-affine we mean that the displacement field is a multi-valued function and therefore $d u_i$ not an exact form, meaning the Burgers vector associated is not zero. More precisely, non affinity means that there exists points in which the infinitesimal displacements cannot be locally approximated by an affine transformation.\\

This brief recap provides concrete proof of all the main three results obtained so far.

\section{The Maxwell viscoelastic model revisited}\label{secnew}
Before moving to defining a formal field theory in terms of the in-in formalism on the SK contour, let us try to discuss what we have found so far, and compare it with the original Maxwell viscoelastic model \cite{maxwell}.\\
The Maxwell model is defined by the following phenomenological (i.e. not based on fundamental symmetries) constitutive relation \cite{maxwell,frenkel}:
\begin{equation}
    \frac{ds}{dt}\,=\,\frac{P}{\eta}\,+\,\frac{1}{G}\,\frac{dP}{dt}
\end{equation}
where $P$ is the shear stress $T_{xy}$ and $s$ the shear strain $\epsilon_{xy}$. In Fourier space, the previous expression becomes:
\begin{equation}
    T_{xy}\,=\,\frac{-\,i\,\omega\,\eta}{1\,-\,i\,\omega\,\tau_M}\,\epsilon_{xy}\,.
\end{equation}
At small frequencies, $\omega \tau_M \ll 1$, we have a pure dissipative viscous response:
\begin{equation}
    T_{xy}\,=\,-\,i\,\omega\,\eta\,\epsilon_{xy}\,,
\end{equation}
while at large frequencies, $\omega \tau_M \gg 1$, a purely elastic one:
\begin{equation}
    T_{xy}\,=\,G\,\epsilon_{xy}\,.
\end{equation}
From here, the idea that Maxwell model interpolates between a fluid behaviour at small frequencies and a solid one at large. In particular, we could interpret the maxwell model as a viscoelastic fluid in which the frequency dependent viscosity reads:
\begin{equation}
    \eta^{\text{Maxwell}}_{\text{eff}}(\omega)\,=\,\frac{\eta}{1\,-\,i\,\omega\,\tau_M}\,.
\end{equation}
In a sense, the k-gap physics is imposed by hands directly in the Maxwell constitutive relation. Notice also that this is very different from the standard hydrodynamic constitutive relation in presence of elasticity and viscosity \cite{PhysRevA.6.2401,chaikin2000principles,Delacretaz:2017zxd,Armas:2019sbe} which reads:
\begin{equation}
     T_{xy}\,=\,\left(G\,-\,i\,\omega\,\eta\right)\,\epsilon_{xy} \label{Kelvin}\,.
\end{equation}
and it is of the Kelvin-Voigt type \cite{10.2307/112142,https://doi.org/10.1002/andp.18922831210} -- summing linearly the various components of the stress.\\

Before introducing phase relaxation, let us explain in detail the difference between the Maxwell model and the Kelvin-Voigt model. The second model describes a material that at zero strain rate ($\omega=0$) is a solid. The Maxwell model on the contrary describes a system that at zero strain rate (or small rate) is a viscous dissipative fluid. By taking the equations of motion for the displacements it is easy to see that the dynamics of collective shear waves is respectively:
\begin{equation}
    \text{Kelvin-Voigt:}\,\,\,\,\,\,\,\,\,\,\,\omega^2\,=\,k^2\,\left(G\,-\,i\,\omega\,\eta\right)
\end{equation}
and:
\begin{equation}
    \text{Maxwell:}\,\,\,\,\,\,\,\,\,\,\,\omega^2\,=\,k^2\,\left(\frac{-\,i\,\omega\,\eta}{1\,-\,i\,\omega\,\tau_M}\right)
\end{equation}
While the Kelvin-Voigt model gives a propagating sound waves with diffusive attenuation, the Maxwell model gives immediately the k-gap dispersion relation.\\
Notice also how the first model is perfectly smooth in the limits $\omega \rightarrow 0$, $\eta \rightarrow 0$ and $G \rightarrow 0$, while the Maxwell mode can become highly singular and the final result dangerously dependent on the order of the limits.\\
Finally, using a Kelvin-Voigt constitutive relation together with phase relaxation, we obtain a frequency dependent viscosity of the type \cite{Delacretaz:2017zxd}:
\begin{equation}
    \eta_{\text{eff}}(\omega)\,=\,\frac{G}{\Omega\,-\,i\,\omega}\,+\,\eta
\end{equation}
There are important differences between the latter and the Maxwell result:
\begin{itemize}
    \item This is not an ad-hoc construction where the final result is basically imposed by hands in the initial phenomenological constitutive relation. Contrarily, it can be formally justified using fundamental symmetry principles.
    \item In both approaches, the frequency dependent viscosity $\eta(\omega)$ displays a Drude peak shape. Nevertheless, in the Maxwell approach, the corresponding relaxation time is forced by hands to be $\tau_M=\eta/G$ without any profound justification. In the second scenario, the relaxation time comes from phase relaxation and it is not a priori related to the shear viscosity $\eta$.
    \item In the Maxwell model the effects are there only at finite (and large) strain rate (i.e. $\omega \neq 0$), while in the case of phase relaxation the effects can be effective also at low frequency.
\end{itemize}
It would be interesting to perform a more detailed comparison between our model and the Maxwell model and confront it with experimental data and simulations. Still, the probably easier way to distinguish the two models is to check carefully whether the relaxation time coincides or not with the Maxwell prediction $\tau = \eta/G$. In this sense, it is also important to provide a formula for the phase relaxation rate $\Omega$ in terms of more microscopic and structural quantities and compare it with the Maxwell one.\\
It is interesting to notice that, in the case of dislocations mediated phase relaxation in solids, the phase relaxation rate can be derived under certain assumptions to be \cite{Delacretaz:2019brr,PhysRevB.22.2514}:
\begin{equation}
    \Omega^{\text{dislocations}}_\perp\,\approx\,n_d\,G\,\frac{\pi\,r_d^2}{2\,\eta_{\text{eff}}}\label{form}
\end{equation}
where $n_d$ and $r_d$ are respectively the density of dislocations and their size and $\eta_{\text{eff}}$ is the effective shear viscosity of the normal state. This implies that, under the assumption $\eta_{\text{eff}}\approx \eta$, we have:
\begin{equation}
    \Omega^{\text{dislocations}}_\perp\,\sim\,\tau_M^{-1}\,.
\end{equation}
It would be extremely important to obtain a formula similar to \eqref{form} for the case of non-affinity induced phase relaxation. A possible relation of this type could potentially provide a final answer to which is the correct relaxation time entering in the k-gap equation \eqref{kgap} and if that has really to do with the microscopic Maxwell time $\tau_M$. It is fair to say that, at the moment, the experimental indications in favour of this last interpretation are very few \cite{doi:10.1063/1.3690083}, if not even absent.

\section{Effective field theory}\label{sec7}
At this point, we want to make a step further, and extend the classical and linear hydrodynamic formulation presented in Section \ref{sec4} to the fully non-linear and out-of-equilibrium regime by using the modern effective field theory methods reviewed in \cite{Glorioso:2018wxw}. We start by constructing the non-equilibrium effective action for solids using the higher-form symmetry picture. Because we have conservation of both lattice momentum and the one-form current (see Section \ref{subsub} for details), we can either choose to work in the standard picture or work in the dual, one-form picture. As a warm-up, we will begin by constructing an effective action in the usual picture, using Goldstone fields corresponding to spontaneously broken lattice momentum. Then we will perform a Legendre transform to find the action describing the dual, one-form picture. In this second, and equivalent, formulation, the fundamental degrees of freedom will not be anymore the standard displacement fields (see also \cite{BEEKMAN20171} for a dual formulation of elasticity in terms of gauge fields). And finally, working in the one-form picture, we will kill the conservation of the higher-form current and see that in the low-frequency regime, transverse waves become diffusive as expected from the presence of phase relaxation and discussed in the previous sections.\\

Solids in which crystal momentum is conserved exhibit the phenomenon of second sound -- thermalized solid phonons act as a fluid through which an independent pressure wave can propagate \cite{PhysRevLett.28.1461,Machida309,lee2019hydrodynamic,doi:10.1002/pssa.2210240102,Cepellotti2015,RevModPhys.46.705}. If crystal momentum is not conserved, then the fluid of thermalized phonons cannot move independently of the lattice and no second sound exists. To keep matters simple, we will neglect the conservation of the stress-energy tensor, which means that we will essentially ignore the fluid degrees of freedom. As a result, if we wish to obtain solid phonons, we must consider theories with conserved lattice momentum. We leave it for future work to incorporate the stress-energy tensor.\\

The actions in the section will be non-equilibrium effective actions defined on the Schwinger-Keldysh (SK) contour. In this way we will be able to account for dissipation. The price we pay is that the field content must be doubled. We will include only the marginal and relevant terms and will satisfy the dynamical  Kubo–Martin–Schwinger (KMS) symmetries \cite{1957JPSJ...12..570K,1959PhRv..115.1342M}. Readers not familiar with these methods should consult \cite{Glorioso:2018wxw,Landry:2019iel}.

\subsection{Solids and their duals}

To begin with, we will construct the quadratic effective action for a solid. At leading-order in the derivative expansion, the non-equilibrium effective action is non-dissipative and can be factorized as the difference of two ordinary actions. Thus, we will work with just one copy of the fields. The field content is 
\begin{equation}\varphi^I(x) = x^I +\phi^I(x) ,\end{equation}
where $\phi^I$ are fluctuations about the equilibrium value of the field. Then, the leading-order quadratic action for an isotropic solid is
\begin{equation}\label{original quadratic} S[\partial_\mu\phi^I] = \int d^3 x \frac{1}{2} \big(\dot{ \vec\phi}^2 - c_{ L }^2 (\vec\nabla \cdot\vec\phi)^2 - c_ T ^2 (\vec\nabla\times \vec\phi)^2 \big) , \end{equation}
which is just the expanded and canonically normalized version of~\eqref{dual action a}. It clearly enjoys shift invariance.
The equations of motion are then 
\begin{equation}\label{quadratic waves 0} \ddot {\vec\phi}_ L  = c_ L ^2 \vec\nabla^2 \vec\phi_ L ,~~~~~~~~~~ \ddot {\vec\phi}_ T  = c_ T ^2 \vec\nabla^2 \vec\phi_ T ,\end{equation}
where we have decomposed $\vec\phi = \vec\phi_ L +\vec\phi_ T $ such that $\vec \nabla\times\vec\phi_ L =\vec\nabla\cdot\vec\phi_ T =0$. We see therefore that $c_ L $ and $c_ T $ are longitudinal and transverse speeds of sound, respectively. \\

We will now perform a Legendre transform to construct a dual action that involves the one-form fields $A_\mu^I$. Begin by replacing $\partial_\mu\phi^I\to V_\mu ^I$; then define the auxiliary action by 
\begin{equation} S_\text{AUX} = S[V_\mu^I]  - \int d^3 x\, \mathfrak{e}^{\mu\nu\lambda}\, V_\mu^I \,F_{\mu\nu}^I, \end{equation} 
where $F^I\equiv d A^I$. Notice that the equations of motion for $A^I$ yield $d V^I=0$,  $V$ is a closed form. Thus, if the spacetime is simply connected, there exists some scalars $\phi^I$ such that $V_\mu^I =\partial_\mu\phi ^I$, $V$ is also exact. Thus, if we integrate out $A^I$ from $S_\text{AUX}$, we obtain the original action~\eqref{original quadratic}. Suppose we instead integrate out $V_\mu^I$. Then we find a local action that only depends on $F^I$. In particular, we have
\begin{equation} \label{dual ordinary action}S_\text{DUAL} = \int d^3 x~ \frac{1}{2} \bigg( -\vec f^2+\frac{1}{c_ L ^2} F_ T ^2 +\frac{1}{c_ T ^2} F_ L ^2 \bigg) , \end{equation}
where 
\begin{equation} f^I = \mathfrak{e}^{ij} \partial_i  A_j^I\,,~~~~~~~~~~F_ T  =\mathfrak{e}^{ij} F_{0 j}^i\,,~~~~~~~~~~F_ L  = F_{0i}^i\,.\end{equation}
We therefore have an action with field content $A^I$, which are the Legendre dual fields of the ordinary phonon fields $\phi^I$. 
The resulting equations of motion are then the constraint equation $\partial_i F_{0i}^I=0$ and the dynamical equations
\begin{equation} \ddot A_ T  = c_ L ^2 \nabla^2 A_ T  ,~~~~~~~~~~ \ddot A_ L  = c_ T ^2 \nabla^2 A_ L , \end{equation}
where $A_ T  = \mathfrak{e}^{ij} A^i_j$, $A_ L  = A^i_i$, and we have gauged-fixed $A_0=0$. Notice that these equations are the dual version of the longitudinal and transverse wave equations~\eqref{quadratic waves 0}.

Finally, let us consider the conserved two-form currents given by
\begin{equation} J^{I\mu\nu} = \frac{\partial S_\text{DUAL} }{\partial F_{\mu\nu}^I} .  \end{equation}
The components are
\begin{equation} \label{one-form current} J^{I ij} = -\mathfrak{e}^{ij} f^I,~~~~~~~~~~ J^{I0i} = \delta^{Ii} \frac{F_ L }{c_ T ^2} +\mathfrak{e}^{Ii}\frac{F_ T }{c_ L^2} .  \end{equation}
Then the equations of motion are equivalent to the conservation equation $\partial_\mu J^{I\mu\nu}=0$. The fact that these two-form currents are conserved ensures that there are no defects in the crystal lattice structure. The nature of this conservation equation is topological and it relates to the vanishing of the Burgers vector $b_i$, implying the absence of non-affine displacements.

\subsection{Amorphous systems and liquids}

To extend the previous action for amorphous systems with non-affine dynamics (liquids and glasses), we must allow defects in the atomic structure to appear and disappear, that is, the one-form charges must not be conserved. This is necessary because of the presence of non-affine displacement fields in these phases. To construct the leading order action with such non-conservation requires the non-equilibrium effective field theory formalism. In this construction, the effective action is defined using the in-in formalism on Schwinger-Keldsyh (SK) contour and therefore has doubled field content $A^I_\mu \to \{ A^I_{1\mu},A^I_{2\mu}\}$~ \cite{kamenev2011field,Landry:2020ire,Landry:2019iel,Landry:2020obv,Glorioso:2018wxw}
. The subscripts 1 and 2 indicate on which leg of the SK contour the fields live. It is convenient to work in the retarded-advanced basis defined by
\begin{equation} A_{r\mu}^I = \frac{1}{2} (A^I_{1\mu}+A^I_{2\mu}),~~~~~~~~~~A^I_{a\mu} = A^I_{1\mu}-A^I_{2\mu} .\end{equation} 
The retarded fields act as classical fields, while the advanced fields encode information about statistical fluctuations. To declutter notation, we will drop the $r$ subscript on the retarded fields. Because the equilibrium state of a glass is thermal, our effective action must be invariant under the dynamical KMS symmetries. Suppose that $\Theta$ is an anti-unitary, time-reversing symmetry of the UV theory and that $\beta_0$ is the inverse equilibrium temperature of the system. Then, the effective action enjoys the symmetries
\begin{equation} A^I_\mu \to \Theta A^I_\mu,~~~~~~~~~~A^I_{a\mu} \to \Theta A^I_{a\mu} + i\Theta \beta_0 \partial_t A^I_\mu. \end{equation}
Finally, there must exist at least one advanced field in each term of the effective action such that terms with even numbers of advanced fields are imaginary and terms with odd numbers of advanced fields are real. The Schwinger-Keldysh EFT version of~\eqref{dual ordinary action} is
\begin{align}
&I_\text{DUAL}[A_{\mu}^I, A_{a\mu}^I] = S_\text{DUAL}[A^I_{1\mu}] -S_\text{DUAL}[A^I_{2\mu}] \nonumber \\&=\int d^3 x \bigg(-\vec f_a \cdot \vec f+\frac{1}{c_ L ^2} F_{a T } F_ T  +\frac{1}{c_ T ^2} F_{a L } F_ L  \bigg). 
\end{align}
Until now, we have simply generalized the (dual) effective action for ordered solids at finite temperature and out-of-equilibrium.\\

If we wish to destroy the conservations of the one-form currents, we must introduce advanced fields without derivatives~\cite{Landry:2020tbh}. It turns out that the correct procedure is to first gauge-fix the one-form fields by working in synchronous gauge, $A_0=A_{a0}=0$~\cite{Landry:2021kko}. Then the residual gauge symmetries are
\begin{equation}A^I_i \to A_i^I + \partial_i \kappa^I(\vec x) , \end{equation}
for time-independent functions $\kappa^I$. Because $\kappa^I$ are time-independent, the SK boundary condition that forces all advanced fields to vanish in the distant future prohibits $A_{ai}^I$ from transforming. It is therefore invariant under the residual gauge symmetries. The leading-order action consistent with these residual gauge symmetries and dynamical KMS is then
\begin{widetext}
\begin{equation}\begin{split} I_\text{NA}[A_{\mu}^I, A_{a\mu}^I] = \int d^3 x \bigg(-\vec f_a \cdot \vec f+\frac{1}{c_ L ^2} F_{a T } F_ T  +\frac{1}{c_ T ^2} F_{a L } F_ L  +\frac{1}{\tau_ L  c_ L ^2} A_{a T } F_ T  +\frac{1}{\tau_ T  c_ T ^2} A_{a L } F_ L  +\frac{i}{\tau_ L  c_ L  \beta_0} A_{a T }^2+\frac{i}{\tau_ T  c_ T ^2 \beta_0} A_{a L }^2 \bigg). \end{split}\label{mio} \end{equation}
\end{widetext}
where the label $NA$ stands for non-affine.\\
Notice that because there are no $\mu=0$ components of $A^I$, we no longer have the constraint equation $\partial_i F_{0i}^I=0$. Instead, we have only the dynamical equations of motion, obtained by varying the action with respect to the advanced fields,
\begin{equation} \ddot A_ T  +\frac{1}{\tau_ L } \dot A_ T = c_ L ^2 \nabla^2 A_ T  ,~~~~~~~~~~ \ddot A_ L +\frac{1}{\tau_ T } \dot A_ L  = c_ T ^2 \nabla^2 A_ L .  \end{equation}
These are our final equations of motion in terms of the new degrees of freedom $A_\mu$. The associated dynamics has been proven above to be equivalent to the standard formulation in terms of the displacement fields $\phi$.
The resulting dispersion relations are then
\begin{equation}\omega^2 + \frac{1}{\tau_s} \omega = c_s^2 k^2,~~~~~~~~~~ s=  L , T  .\end{equation}
In the low-frequency limit $\omega\tau \ll 1$, the dispersion relations become $\omega =-i D_s k^2$ for $D_s\equiv \tau_s c_s^2$. By contrast in the high-frequency limit $\omega \tau \gg 1$, the dispersion relations become $\omega = c_s k$. We therefore see that there are two qualitatively different behaviors depending on the frequency: dispersion in the IR and propagating waves in the UV. 

Finally, let us consider the one-form current~\eqref{one-form current}. Unlike in the previous example, this current is now no longer conserved, in particular, we have
\begin{equation} \partial_\mu J^{\mu }_ T  = -\frac{1}{\tau_ T } J^t_ T , ~~~~~~~~~~\partial_\mu J^{\mu }_ L  = -\frac{1}{\tau_ L } J^t_ L , \label{bre}\end{equation}
where $J_ T ^\mu \equiv J^{i \mu i}$ and $J_ L ^\mu \equiv \mathfrak{e}^{ij}J^{i \mu j}.$ Now that these currents are no longer conserved, the crystal lattices may posses defects. Indeed, defects may arise and disappear freely. Notice that we now have no conserved quantities and hence no propagating waves as $\omega\to 0$. What happened to the hydrodynamic phonon? Recall that at the beginning, we neglected the conservation of the stress-energy tensor for he sake of simplicity. If we were to include its conservation, then the hydrodynamic sound mode would be restored. \\
Notice that, if one would attempt from the action \eqref{mio} to go back to the original formulation in terms of the Goldstone fields, one would get a non-local, and therefore not tractable, action. In this sense, this dual picture is extremely convenient to describe from the field theory point of view the effects of phase relaxation.\\

Finally, let us remind the Reader that the breaking of the conservation equations in \eqref{bre} is equivalent to the presence of a finite Burgers vector (due to non-affine dynamics) and of a non-trivial topological structure.

\section{A topological scenario for glasses}\label{secglass}
One of the biggest unsolved problems in modern physics is clarifying the nature of the glass transition. In particular, there is an apparent contradiction between the impossibility to tell apart glasses and liquids at the level of two-point correlation functions, such as the radial distribution function $g(r)$ -- which looks identical for the two states --, and the huge difference between liquid and glass in terms of rigidity and mechanical properties (although confined liquids have been shown to behave solid-like under certain boundary conditions~\cite{Zaccone_PNAS_2020,Noirez_2012}).
Traditionally~\cite{Berthier_review}, two different views can be traced back to whether a true phase transition separates liquid and glass (i.e. the ideal glass transition, supposed to exist at some unattainable low temperature)~\cite{Thirumalai} or the two states are merely separated by a dynamical crossover~\cite{Goetze}. In other words, it is an open question whether glasses and liquids are fundamentally different (in terms of symmetries for example) or simply the manifestation of the same kind of physics but with a very different characteristic timescale.

\begin{figure}[ht]
    \centering
    \includegraphics[width=\linewidth]{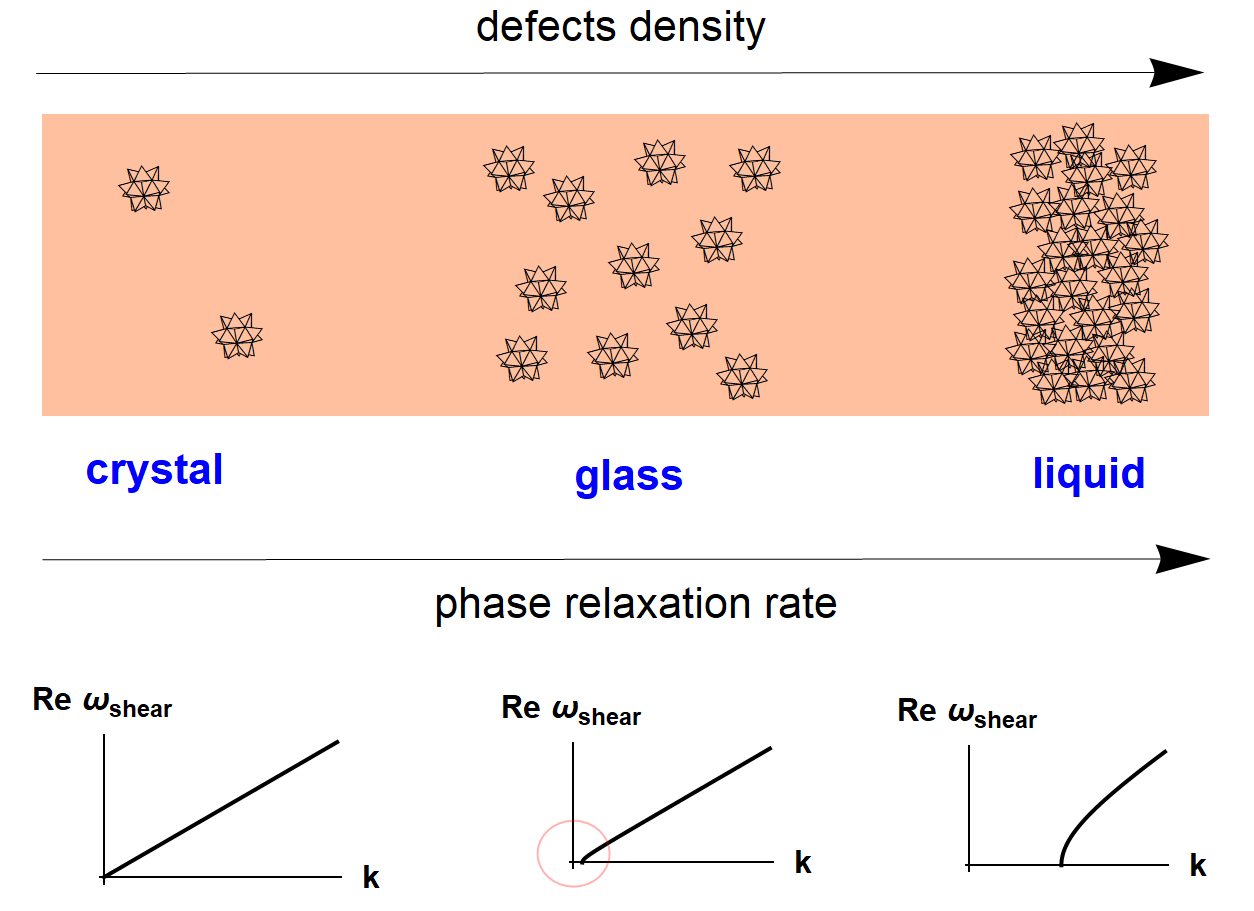}
    \caption{The correlation between non-affine dynamics, topological defects density, phase relaxation rate $\Omega$, and the gap of the collective shear waves in ordered crystals, glasses and liquids. By topological defects we simply mean non-affine displacements with a non-zero Burgers vector.}
    \label{fig:glass}
\end{figure}

The overall picture that emerges in the previous sections of this work, suggests this second option. Well-defined topological defects in amorphous systems can be identified, not in the static structure of liquids or glasses, but instead in the displacement field $u_{i}$ of the material under deformation. In particular, well-defined topological defects can be associated with the non-affine displacement field, as demonstrated above.
The key finding of this work, that the phase relaxation rate $\Omega$, which is responsible for killing the Goldstone phonons at low frequency, correlates positively with the non-affinity of the displacement field under small deformations, suggests that the Burgers vector associated with the non-affine field grows upon going from glass to liquid, i.e. upon increasing the temperature
~\cite{Lemaitre_PRL_2013,Wittmer}, or in the case of strain-induced glass-liquid transition, upon increasing the strain magnitude~\cite{Schall}.
In particular, growing non-affine displacements lead to growing $\Omega$, which in turn broadens the k-gap of the shear phonons. Conversely, upon decreasing the temperature $T$ in a supercooled liquid, leads to lower values of $\Omega$, and therefore, to a lower extent of topological two-form symmetry breaking, and to a narrower k-gap in the transverse sector, as schematically shown in Fig.~\ref{fig:glass}. This eventually results in a vanishingly small k-gap that remains frozen-in at the glass transition, and provides an apparent shear rigidity, with $G>0$ down to low-frequency/rate of deformation. This scenario of a topologically-driven crossover controlled by the non-affine displacement field, is compatible with the dynamical crossover view of the glass transition, and with the substantial continuity between liquid and solid state across the glass transition first proposed by Frenkel~\cite{Frenkel1935}. The ``apparent'' shear elasticity is also consistent with recent claims about the non-existence of a true shear continuum elasticity in glasses~\cite{Biroli2016}.  Our analysis clarifies also that the Goldstone modes of glasses in the transverse sector are diffusive because of the presence of phase relaxation, and not because of the breaking of momentum conservation as claimed in \cite{Vogel_2019}. Clearly, the breaking of the transverse momentum conservation would correspond to an explicit breaking of translations as in the Drude model, and it would destroy any hydrodynamic modes in the transverse sector by introducing necessarily a finite damping term.

Instead, the transition from liquid to an isotropic crystalline solid is a true phase transition accompanied by the divergence of the phase relaxation time $1/\Omega$, since $\Omega = 0$ in the perfect crystalline phase (and $\Omega \approx 0$ when only very few tolopogical defects are present). In this case, there is no gap in the dispersion relations for the transverse phonons of the solid, which reach all the way down to $k=0$. 

\section{Summary}\label{sec8}
In this work, we have developed a formal and self-consistent theory of solids, liquids and glasses based on phonons as Goldstone bosons and phase relaxation related to topological input due to non-affine deformations. Thanks to the interplay of these two mechanisms, this formalism is able to encompass all three different phases of matter and to provide a deeper understanding in terms of fundamental symmetries and topological defects.

Using different theoretical methods, we have been able to connect the existence of microscopic non-affine displacements in liquids and glasses with the breaking of a topological-like two-form symmetry. Moreover, we have shown the equivalence between the latter, the compatibility equation for the strain tensor, the presence of a non-trivial Burgers vector for non-affine deformations, and the macroscopic mechanism of phase relaxation for the Goldstone modes, the phonons.

As a direct application of our formalism, we have derived the appearance of propagating shear waves in liquids beyond  a critical momentum cutoff, known as k-gap \cite{BAGGIOLI20201}. This is a more fundamental and symmetries-compatible derivation of the k-gap in liquids, which improves on the phenomenological approaches of Maxwell \cite{maxwell}, Frenkel \cite{frenkel}, and later Trachenko and collaborators \cite{Trachenko_2015}, whose physical arguments were certainly pioneering and valid, but lacking a rigorous formal background. As a matter of fact, the simple Maxwell model is unable to correctly reproduce recent atomistic simulations data of frequency-dependent viscoelastic moduli of simple liquids~\cite{Sader}.
Moreover, our approach could shed new light on the problem of the glass transition, at which the phase relaxation time becomes very large (though not infinite as in solids), and on the nature of the Frenkel line, at which the phase relaxation time saturates the Debye time-scale (see also recent conjectures \cite{2021arXiv210103983C}).\\

\begin{figure}
    \centering
    \includegraphics[width=\linewidth]{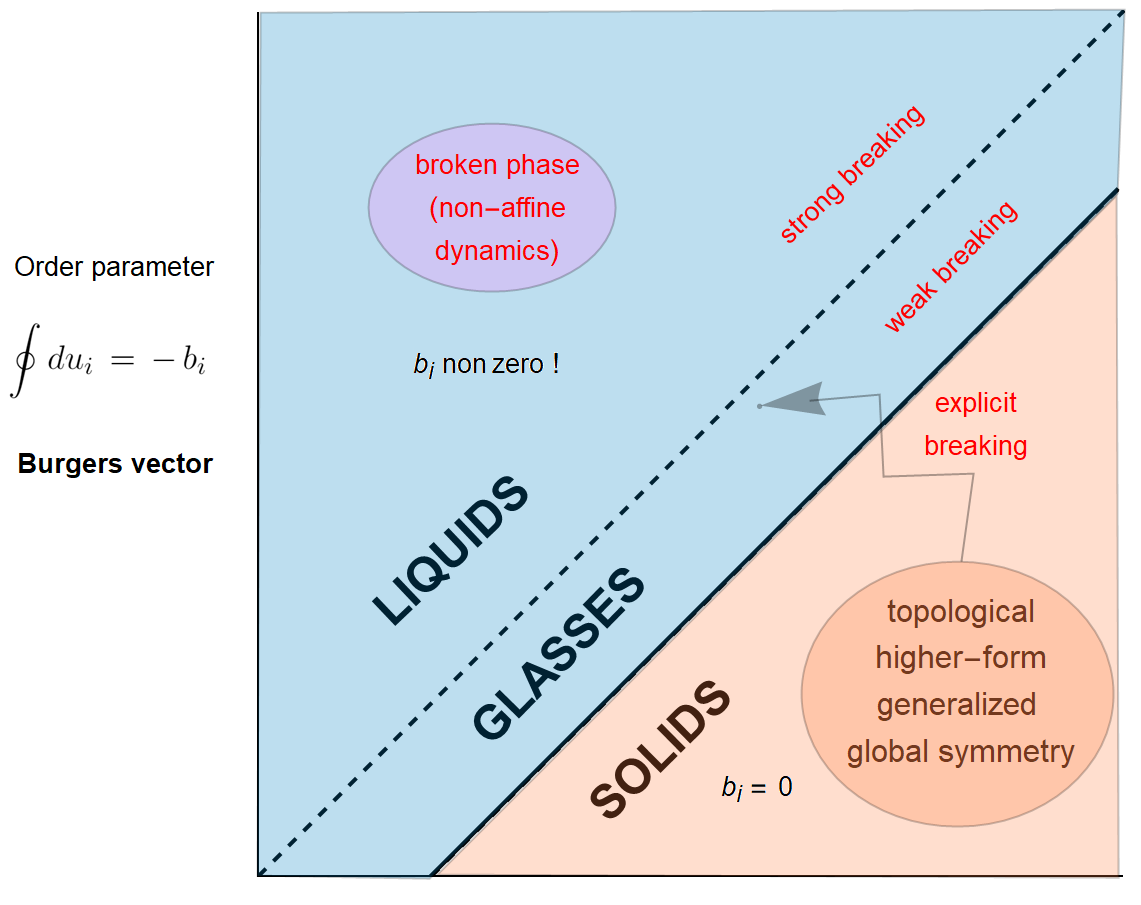}
    \caption{The phase diagram in terms of the topological higher-form generalized global symmetry, which corresponds to the conservation of the two-form $\partial_\mu J^{\mu\nu}_I=0$, and of the topological order parameter -- the Burgers vector $b_i$. The filled line indicates the explicit breaking of such a symmetry due to the appearance of non-affine displacements. It also distinguishes the phases with zero Burgers vector $b_i=0$ (solids) from the ones (glasses, fluids) with a finite value for this topological order parameter. The dashed line between liquids and glasses cannot be defined by any fundamental symmetry, but only by the strength of the symmetry breaking, weak for glasses (corresponding to a very long relaxation time) and strong for liquids. In this view, the difference between solids and amorphous systems (liquids and glasses) is purely topological and not linked to any spacetime symmetry.}
    \label{phased}
\end{figure}

From a formal point of view, our theory demonstrates that liquids, solids and glasses do not present any fundamental difference at the level of spacetime symmetries. This conclusion is a unification of the original ideas of Leutwyler for solids \cite{Leutwyler:1996er} and of those of Frenkel in liquids \cite{frenkel}. Nevertheless, we prove that the distinction between ordered solids and amorphous systems has to be identified with the broken (or not) generalized higher-form global symmetries, which is the only fundamental ingredient able to provide an accurate phase diagram based on symmetries (see Fig.~\ref{phased}). In this language, a solid is a phase where such a symmetry is nonlinearly realized, while an amorphous system (glasses and fluids) is a phase of matter where the higher-form symmetry is explicitly broken because of the underlying non-affine dynamics. As such, the distinction between solids and fluids is topological in nature and intrinsically connected to the geometrical and dynamical nature of the deformation field therein.
Additionally, the Burgers vector associated to the dynamical displacement fields serves as a topological order parameter distinguishing those phases. It is indeed zero in ordered solids with no defects but inherently finite in liquids and glasses.
In a companion paper \cite{burgersking}, we show the application of these ideas to glasses, proving from numerical data the existence of a finite Burgers vector associated with the phase relaxation mechanism discussed in this paper.\\

Several are the ideas to pursue in the future:
\begin{itemize}
    \item First, our results unveil the importance of topological effects in amorphous systems, by providing a working definition of topological defects in the displacement field of glasses and liquids. This has recently been advocated in various directions \cite{PhysRevLett.125.118002,PhysRevLett.118.236402,grushin2020topological} and it certainly is an interesting point to expand upon further.
    \item It would be very helpful to use a microscopic model for non-affine displacements (e.g. \cite{Zaccone_2011}) to compute directly the phase relaxation rate, and compare it with (a) the Maxwell prediction and (b) the (few) experimental data available.
    \item Recently, a formal duality between elasticity with topological defects and \textit{fractons} has been shown \cite{PhysRevB.100.134113}. In this picture, the dynamics of dislocations and disclinations is viewed in terms of dynamical fractonic degrees of freedom with reduced mobility. It would be interesting to establish if this equivalence could be useful to understand the role of fractons for liquids dynamics. A first connection between higher-form symmetries, their breaking and fractonic phase has been initiated in \cite{Qi:2020jrf}.
    \item From a theoretical perspective, it would be nice to see if phase relaxation could be re-written as a quantum anomaly of the higher-form global symmetry in the spirit of \cite{Delacretaz:2019brr,Landry:2021kko}.
    \item Recently, a field theory for amorphous solids has been proposed in \cite{PhysRevLett.121.118001} based on previous works \cite{PhysRevLett.95.198002}. It would be interesting to find possible connections between our work and that formalism.
    \item The physics of phonons in liquids share several features with the idea of electromagnetic screening and even confinement. Some basic analogies have been discussed in \cite{BAGGIOLI20201}. It would be exciting to continue on these lines by using the connection between higher-form symmetries, spontaneous symmetry breaking and Wilson loops \cite{Hofman:2018lfz}.
    \item The same physics described in this manuscript via the presence of phase relaxation appears in the framework of Hydro+ \cite{Stephanov:2017ghc} by considering the effects of hydrodynamic fluctuations and the presence of slowly relaxing additional modes. It would be very interesting to draw a more explicit connection between these concepts. Can non-affinity and phase relaxation be understood in terms of hydrodynamic fluctuations?
    \item The disappearance of the Goldstone modes at low frequencies in glasses, especially in relation to the sample size $L$, is a hot topic of discussions. Several approaches have been proposed \cite{2015arXiv150200685W,PhysRevLett.115.267205,C3SM50998B,gartner2016nonlinear}, but none of them related to any fundamental symmetry or symmetry-breaking. Here, we have provided a first-principle derivation of the appearance/disappearance of transverse Goldstone modes in glasses and liquids based on the topological character of non-affine displacements and phase relaxation. It would be very interesting to study the system size dependence of the phase relaxation rate and see if our framework could explain the findings of \cite{gartner2016nonlinear} and complement recent research on finite-size effects of elasticity of liquids and glasses~\cite{Zaccone_PNAS_2020,Noirez_2012,Noirez2021,phillips2020universal}. This would be achieved by a phase relaxation rate $\Omega$ decreasing with the system size $L$. Also, are the ``glassy quasilocalized
   excitations''~\cite{Lerner} perhaps related the low frequency diffusive remnants of the transverse Goldstone modes discussed in this work? In our language, the crossover between the quasilocalized modes and the propagating standard Goldstone modes is controlled by the phase relaxation rate and the density of non-affine displacements. The proliferation of non-affine displacements processes is the responsible for the appearance of these quasilocalized modes at frequencies lower than a certain scale controlled by the phase relaxation rate $\Omega$.
\item In this manuscript, we have presented the idea that the Burgers vector (or more precisely its norm) could serve as an appropriate topological order parameter distinguishing liquids (and glasses) from solids. It would be important to complete this picture and determining the behaviour of such order parameter across the liquid-solid phase transition (suggested, in the above, to be a topological transition).
    \item An ultimate and very ambitious goal would be to build a symmetries-based effective field theory for glasses able to predict and describe all their well-known anomalous properties.
\end{itemize}
We leave these questions for the future and for the interested Reader.

\subsection*{Acknowledgments} 
We thank Saso Grozdanov, Maria Vozmediano, Blaise Gouteraux, Vincenzo Vitagliano, David Pereniguez Rodriguez, Nick Poovuttikul and especially Kostya Trachenko for useful discussions and fruitful comments. A.Z. acknowledges financial support from US Army Research Office, contract nr. W911NF-19-2-0055. M.B. acknowledges the support of the Shanghai Municipal Science and Technology Major Project (Grant No.2019SHZDZX01) and
of the Spanish MINECO “Centro de Excelencia Severo Ochoa” Programme under grant
SEV-2012-0249. M.L. thanks Alberto Nicolis and Lam Hui for their mentorship and guidance and acknowledges financial support from US Department of Energy grant DE-SC011941. 
\bibliographystyle{apsrev4-1}

\bibliography{biblio}

\end{document}